\documentclass{cta-author}
\usepackage{multirow}
\usepackage[font=scriptsize,caption=false,labelsep=space]{subfig}
\usepackage{mathrsfs}
\usepackage{bm}
\usepackage{graphicx,float,wrapfig,epstopdf,amsmath}
\usepackage[]{algorithmicx}
\usepackage{algpseudocode,algorithm}
\usepackage{balance}
{}
{}
{}
{}
{}

\begin{document}
\setlength{\abovedisplayskip}{3pt}
\setlength{\belowdisplayskip}{3pt}

	\supertitle{Research Article}
	
	\title{Cognitive Radar Antenna Selection via Deep Learning}
	
	\author{\au{Ahmet M. Elbir$^{1}$}, \au{Kumar Vijay Mishra$^{2\corr}$}, \au{Yonina~C. Eldar$^{2}$}}
	
	\address{\add{1}{Department of Electrical and Electronics Engineering, Duzce University, Duzce, Turkey}
		\add{2}{Andrew and Erna Viterbi Faculty of Electrical Engineering, Technion-Israel Institute of Technology, Haifa, Israel}
		\email{mishra@ee.technion.ac.il}}
	
	\begin{abstract}
    	Direction of arrival (DoA) estimation of targets improves with the number of elements employed by a phased array radar antenna. Since larger arrays have high associated cost, area and computational load, there is recent interest in thinning the antenna arrays without loss of far-field DoA accuracy. In this context, a cognitive radar may deploy a full array and then select an optimal subarray to transmit and receive the signals in response to changes in the target environment. Prior works have used optimization and greedy search methods to pick the best subarrays cognitively. In this paper, we leverage deep learning to address the antenna selection problem. Specifically, we construct a convolutional neural network (CNN) as a multi-class classification framework where each class designates a different subarray. The proposed network determines a new array every time data is received by the radar, thereby making antenna selection a cognitive operation. Our numerical experiments show that {the proposed CNN structure provides 22\% better classification performance than a Support Vector Machine and the resulting subarrays yield 72\% more accurate DoA estimates than random array selections.} 
	\end{abstract}
	
	\maketitle
	\section{Introduction}
	Cognitive radar has gained much attention in the last decade due to its ability to adapt both the transmitter and receiver to changes in the environment and provide flexibility for different scenarios as compared to conventional radar systems \cite{cognitiveRadarHaykin, cognitiveRadarRef2, cognitiveRadarExperiments}. Several applications have been considered for radar cognition such as waveform design \cite{cognitiveRadarWaveformDesign, cognitiveRadarWaveformDesignReceiverSelection,mishra2017performance,mishra2018cognitive}, target detection and tracking \cite{cognitiveRadarTracking,cognitiveRadarRecognition}, spectrum sensing and sharing \cite{cognitiveRadarSpectrumSharing,cohen2017spectrum,mishra2018sub,cohen2018sub,na2018tendsur}. Cognitive radar design requires reconfigurable circuitry for many subsystems such as power amplifiers, waveform generator, and antenna arrays \cite{cognitiveRadarReconfigurableCircuitry}. In this paper, we focus on this latter aspect of antenna array design in cognitive radar.
	
    For a given wavelength, good angular resolution is achieved by a wide array aperture resulting in a large number of array elements, physical area and the cost associated with the array circuitry \cite{RPA2,antennaSelectionForMIMO,ref_AI4}. Hence, general approaches have been proposed to effectively use the array output with minimal number of antenna elements. For example, non-uniform array structures \cite{coprimeDSPConf, nestedArray} are used to virtually increase the array aperture for direction-of-arrival (DoA) estimation. Given a full Nyquist antenna array, one could also randomly choose a few antenna elements to transmit/receive (Tx/Rx) and then employ efficient recovery algorithms so that the spatial resolution does not degrade \cite{thinnedArray,RPA2,rossi2014spatial,suMMeRPpaper,mishra2018cognitive}. However, such approaches are agnostic to information about the received signal. A cognitive approach may be to select these Tx/Rx antennas based on the current target scenario encoded in the received signal \cite{cognitiveRadarReconfigurableCircuitry} connecting antenna selection with contemporary interest in cognitive radar. The key idea is to exploit available data from the current radar scan to choose an optimal subarray for the next scan since target locations change little during consecutive scans.
    
   Recent research \cite{antennaSelectionCognitive, antennaSelectionCognitive2, mateos2017adaptive} has proposed a reconfigurable array structure for a cognitive radar which obtains an adaptive switching matrix after a combinatorial search for an optimal subarray that minimizes a lower bound on the DoA estimation error. A related work in \cite{antennaSelectionMultipleWavelengthSensing} proposed a greedy search algorithm to find a subarray that maximizes the mutual information between the collected measurements and the far-field array pattern. Very recently, a semidefinite program proposed in \cite{antennaSelectionRxTxPairSelection} selects a Tx-Rx antenna pair for a multiple-input-multiple-output (MIMO) radar that maximizes the separation between desired and parasitic DoAs. Similar problems have also been investigated in communications especially in the context of massive MIMO \cite{commPaperMassiveMIMO} to achieve energy and cost efficient antenna designs and beamforming \cite{antennaSelectionTxMIMO,antennaSelectionMISO,antennaSelectionMulticasting}. More generally, in the context of sensor selection, \cite{antennaSelectionViaCO,sparsityEnforcingSS} solve convex optimization problems to obtain optimal antenna subarrays for DoA estimation. Similarly, \cite{antennaSelectionKnapsack} selects the sensors for a distributed multiple-radar scenario through greedy search with the Cram\'{e}r-Rao lower bound (CRB) as a performance metric. 
    
	Nearly all of these formulations solve a mathematical optimization problem or use a greedy search algorithm. A few other works explore supervised machine learning (ML) to estimate DoA in the context of radar \cite{machineLearningRadarDetection} and communications \cite{svmDoAEst1, svmDoAEst2}. Specifically, \cite{machineLearningAntennaSelection} employs support vector machines (SVM) for antenna selection in wireless communications. As a class of ML methods, deep learning (DL) has gained much interest recently for the solution of many challenging problems such as speech recognition, visual object recognition, and language processing \cite{deepLearningScience, deppLearningRepresetation}. DL has several advantages such as low computational complexity when solving optimization-based or combinatorial search problems and the ability to extrapolate new features from a limited set of features contained in a training set \cite{svmDoAEst1,deepLearningScience}. In the context of radar, DL has found applications in waveform recognition \cite{deepLearningRadarRecognition}, image classification \cite{deepLearning4Radar,deepLearningSAR}, range-Doppler signature detection \cite{deepLearnnigRangeDopplerRadar}, and rainfall estimation \cite{mishra2018deep}. 
	 
	In this paper, we introduce a DL-based approach for antenna selection in a cognitive radar. DL techniques directly fit our setting because the antenna selection problem can be considered as a classification problem where each subarray designates a class. Among prior studies, the closest to our work is \cite{machineLearningAntennaSelection} where SVM is fed with the channel state information (CSI) to select subarrays for the best MIMO communication performance. However, SVM is not as powerful as DL for extracting feature information inherit in the input data \cite{deepLearningScience}. Furthermore, \cite{machineLearningAntennaSelection} considers only small array sizes. On the other hand, the optimization methods suggested in \cite{antennaSelectionKnapsack,sparsityEnforcingSS} assume \textit{a priori} knowledge of the target location/DoA angle to compute the CRB. Compared to these studies, we leverage DL to consider a relatively large scale of the selection problem wherein the feature maps can be extracted to train the network for different array geometries. The proposed approach avoids solving a difficult optimization problem \cite{antennaSelectionViaCO}. Unlike random array thinning where a fixed subarray is used for all scans, we select a new subarray based on the received data. In contrast to \cite{antennaSelectionKnapsack,sparsityEnforcingSS}, we also assume that the target DoA angle is unknown while choosing the array elements. To the best of our knowledge, this is the first work that addresses the radar antenna selection problem using DL. 
    
     In particular, we construct a convolutional neural network (CNN) for our problem. The input data to our CNN are the covariance samples of the received array signal. Previous radar DL applications \cite{deepLearningRadarRecognition, deepLearning4Radar, deepLearningSAR, deepLearnnigRangeDopplerRadar} have used image-like inputs such as synthetic aperture radar signatures and time-frequency spectrograms. Our proposed CNN models the selection of $K$ best antennas out of $M$ as a classification problem wherein each class denotes an antenna subarray. In order to create the training data, we choose those subarrays which estimate DoA with the lowest minimal bound on the mean-squared-error (MSE). We consider minimization of CRB as the performance benchmark in generating training sets for 1-D uniform linear arrays (ULA) and 2-D geometries such as uniform circular arrays (UCA) and randomly deployed arrays (RDA). For ULAs, we also train the network with data obtained by minimizing Bayesian bounds such as the Bobrovsky-Zakai bound (BZB) and Weiss-Weinstein bound (WWB) on DoA \cite{performanceBoundsWWB} because these bounds provide better estimates of MSE at low signal-to-noise-ratios (SNRs). In particular BZB-based selection has been shown to have the ability to control the sidelobes and avoid ambiguity in DoA estimation \cite{antennaSelectionCognitive,antennaSelectionCognitive2}. {Our results show that the proposed CNN classification performance is $22$\% better than the SVM. The DoA estimation accuracy using the subarrays obtained from our CNN network is $32$\% and $72$\% higher than the SVM and random array selections, respectively.}
	
	The rest of the paper is organized as follows. In the next section, we describe the system model and formulate the antenna selection problem. In Section \ref{sec:DNN}, we introduce the proposed CNN and provide details on the training data. We evaluate the performance of our DL method in Section \ref{sec:Simulations} through several numerical experiments.
    
	Throughout the paper, we reserve boldface lowercase and uppercase letters for vectors and matrices, respectively. The $i$th element of a vector $\textbf{y}$ is $y(i)$ while the $(i,j)$th entry of the matrix $\textbf{Y}$ is $[\textbf{Y}]_{i,j}$. We denote the transpose and Hermitian by $(\cdot)^T$ and $(\cdot)^H$, respectively. The functions $\angle\left\lbrace \cdot \right\rbrace$, $\operatorname{\mathbb{R}e}\left\lbrace \cdot \right\rbrace$ and $\operatorname{\mathbb{I}m}\left\lbrace \cdot \right\rbrace$ designate the phase, real and imaginary parts of a complex argument, respectively. The combination of selecting $K$ terms out of $M$ is denoted by $\left( \begin{array}{c} M \\ K \end{array}\right) = \frac{M !}{K!(M-K)!}$. The calligraphic letters $\mathcal{S}$ and $\mathcal{L}$ denote the position sets of all and selected subarrays, respectively. The Hadamard (point-wise) product is written as $\odot$. The functions $\text{E}\left\lbrace \cdot \right\rbrace $ and $\text{max}$ give the statistical expectation and maximum value of the argument, respectively. The notation $x \sim \textrm{u}\{[u_l,u_u]\}$ means a random variable drawn from the uniform distribution over $[u_l,u_u]$ and $x \sim \mathcal{CN}(\mu_x,\sigma_x^2)$ represents the complex normal distribution with mean $\mu_x$ and variance $\sigma_x^2$.
	\vspace{-12pt}
	\section{System Model and Problem Formulation}
	\label{sec:SystemModel}
    	Consider a phased array antenna with $M$ elements where each element transmits a pulsed waveform $s(t)$ towards a Swerling Case 1 point target. Since we are interested only in DoA recovery, the range and Doppler measurements are not considered and target's complex reflectivity is set to unity. {Assumption of a Swerling I model implies that the target parameters remain constant for the duration of the scan}. We characterize the target through its DoA $\Theta=(\theta,\phi)$ where $\theta$ and $\phi$ denote, respectively, the elevation and the azimuth angles with respect to the radar. The radar's pulse repetition interval and operating wavelength are, respectively, $T_s$ and $\lambda= c_0/f_0$, where $c_0 = 3\times10^8$ ms$^{-1}$ is the speed of light and $f_0=\omega_0/2\pi$ is the carrier frequency.
        
         To further simplify the geometries, we suppose that the targets are far enough from the radar so that the received signal wavefronts are effectively planar over the array. The array receives a narrowband signal reflected from a target located in the far-field of the array at $\Theta$. We denote the position vector of the $m$th receive antenna by $\textbf{p}_m = [p_{x_m},p_{y_m},p_{z_m}]^T$ {in a Cartesian coordinate system} and 
assume that the antennas are identical and well calibrated.
         
         Let $s(t)$ and $y_m(t)$ denote the source signal and the output signal at the $m$th sensor of the array, respectively. The baseband continuous-time received signal at the $m$th antenna is then
	\begin{align}
	\label{sigModelInTime}
    y_m(t) = a_m(\Theta)s(t) + n_m(t),\phantom{1}\phantom{1}0 \le t \le T_s,
    \end{align}    
	where $n_m(t)$ is temporally and spatially white zero-mean Gaussian noise with variance $\sigma_n^2$ and 
    \begin{align}
    \label{eq:stevec1}
	{a_m(\Theta) = \exp \left\{j\frac{2\pi}{\lambda} c_0\tau_m \right\}},
	\end{align}
    is the $m$th element of the steering vector 
    \begin{align}
    \label{eq:stevec2}
    \textbf{a}(\Theta)=[a_1(\Theta),a_2(\Theta),\dots, a_M(\Theta)]^T.
    \end{align}
   Here, $\tau_m$ is the time delay from the target to the $m$th antenna {with respect to the reference antenna in the array}, and is given by $\tau_m = -\frac{1}{c_0}\textbf{p}_m^T\textbf{r}(\Theta)$ where $\textbf{r}(\Theta)$ is the 2-D \textit{DoA parameter}
	\begin{align}
	\textbf{r}(\Theta) = [\cos(\phi)\sin(\theta), \sin(\phi)\sin(\theta), \cos(\theta)]^T.
	\end{align}
    
    The radar acquires the signal {for the $l$th snapshot over a pulse repetition interval (PRI) $T_a$. For a given snapshot $l$, we define an $M\times 1$ received signal vector $\textbf{y}(lT_a) = [y_1(lT_a),\dots,y_M(lT_a)]^T$. For all $L$ snapshots, omitting $T_a$} from the indices for notational simplicity, we can express the received signal in matrix form as
	\begin{eqnarray}
	\textbf{Y} = \textbf{a}(\Theta)\textbf{s}^T + \textbf{N},
    \end{eqnarray}
    where $\textbf{Y}$ is the $M\times L$ matrix given as $\textbf{Y} = [\textbf{y}(1),\cdots,\textbf{y}(L)]$, $\textbf{s}=[s(1),\dots,s(L)]^T$, and $\textbf{N}=[\textbf{n}(1),\cdots,\textbf{n}(L)]$ with $\textbf{n}(l) = [n_1(l),\dots,n_M(l)]^T$ denoting the noise term.

Expanding the inner product $\textbf{p}_m^T \textbf{r}(\Theta)$ in the array steering vector gives $a_m(\Theta) = \exp\left\{{-}j\dfrac{2\pi}{\lambda} (p_{x_m} \mu + p_{y_m} \nu + p_{z_m}\xi)\right\}$, where $\mu = \cos(\phi)\sin(\theta)$, $\nu = \sin(\phi)\sin(\theta)$, and $\xi=\cos(\theta)$. Evidently, $a_m(\Theta)$ is a multi-dimensional harmonic. Once the frequencies $\mu$, $\nu$, and $\xi$ in different directions are estimated, the DoA angles are obtained using the relations 
\begin{align}
\label{eq:thetaphiEst}
\theta = \tan^{-1}\left(\frac{\nu}{\mu}\right),\;\;\; \phi=\cos^{-1}\left(\frac{\xi}{\sqrt{\mu^2+\nu^2+\xi^2}}\right),
\end{align}
with the usual ambiguity in $[0, 2\pi]$. In case of a linear array, there is only one parameter in the steering vector whereas two parameters are involved in planar and three-dimensional arrays.
    
For instance, consider a planar array so that $p_{z_m}=0$ and there is only one incoming wave. Then, a minimal configuration to find the two frequencies consists of at least three elements in an L-shape configuration to estimate frequencies in the $x-$ and $y-$directions. More sensors are needed if the incoming signal is a superposition of $P$ wavefronts. Many theoretical works have investigated the uniqueness of 2-D harmonic retrieval (see e.g., \cite{jiang2001almost,liu2002constant,nion2010tensor}). For example, in case of a uniform rectangular array of size $M_1\times M_2$, classically $P \le M_1M_2 - \text{min}(M_1,M_2)$ specifies the minimum required number of sensors in the absence of noise. This can be relaxed by obtaining several snapshots or using coprime sampling if the sources are uncorrelated \cite{coprimeDSPConf,nestedArray} or spatial compressed sensing \cite{thinnedArray,rossi2014spatial}.

In the antenna selection scenario, the radar has $M$ antennas but desires to use only $K<M$ elements to save computational cost, energy and sharing aperture momentarily to look at other directions. In practice, the signal is corrupted with noise and the antenna elements are not ideally isotropic. Therefore, $K$ should be larger than the minimum number elements predicted by classical results. Removal of elements from an array raises the sidelobe levels and introduces ambiguity in resolving DoAs. The exact choice of $K$ depends on the estimation algorithm employed by the receiver processor. For example, \cite{nion2010tensor,thinnedArray,rossi2014spatial} provide different guarantees for the minimum $K$ depending on the array configuration and algorithm used for extracting the DoA.

In general, the target's position changes little during consecutive scans while a phased array can switch very fast from one antenna configuration to the other. {Here, we consider the following scan strategy for the radar: at the beginning (the very first scan), all $M$ antennas are active and the received signal from this scan is fed to the network. Our goal is to find an optimal antenna array for the next scan in which only $K$ antennas will be used. The radar continues to use this subarray for a few subsequent scans. After surveying the target scene with this optimal subarray for a predetermined number of scans, the radar switches back to the full array for a single scan. The received signal from this full array scan is then used to find a new, optimal subarray for the subsequent few scans. The frequency of choosing a new subarray can be decided off-line based on the nature of the target and analysis of previous observations.} This switching of elements between different scans is a cognitive operation because a new array is determined in every few scans based on received echoes from the target scene. In the next section, we describe our DL approach to cognitive antenna selection. 
	\vspace{-12pt}
	\section{Deep Neural Network for Antenna Selection}
	\label{sec:DNN}
	Assume that an antenna subarray composed of $K$ antennas is to be selected from an $M$-element antenna array. There are $Q = \left( \begin{array}{c} M \\ K \end{array}\right) $ possible choices. This can be viewed as a classification problem with $Q$ classes each of which represents a different subarray. Let $\mathcal{P}_k^q = \{p_{x_k}^q,p_{y_k}^q,p_{z_k}^q\}$, $k=1,\dots,K$, be the set of the $k$th antenna coordinates in the $q$th subarray. Then, the $q$th class consisting of the positions of all elements in the $q$th subarray is
	\begin{align}
	\mathcal{S}_q =  \{\mathcal{P}_1^q,\dots,\mathcal{P}_K^q  \},
	\end{align}
    and all classes are given by the set
	\begin{align}
	\mathcal{S} = \{\mathcal{S}_1,\mathcal{S}_2,\dots, \mathcal{S}_Q  \}.
	\end{align}
    
   In Section~\ref{subsec:network}, we propose a CNN to solve this classification problem {by selecting the positions of the antenna subarray that provides the best DoA estimation performance.} We discuss in Section~\ref{subsec:inputDataGeneration} our procedure to create training data relying on the CRB. Note that, for an operational radar, generation of an artificial or simulated dataset to train the DL network is not necessary. Instead, the network can train itself with the data acquired by the radar during previous scans. During the test phase, for which we present results in Section~\ref{sec:Simulations}, the DoA angles are unknown to the network. The CNN accepts the features from the estimated covariance matrix and outputs a new array. This stage is, therefore, cognitive because the radar is adapting the antenna array in response to the received signal. {In our simulations, we observe that the proposed network can provide robust antenna selection performance with up to 2 degrees DOA mismatch between the data of consecutive scans, where a UCA is considered for $M=20$, $K=5$ and SNR $=15$ dB.} 
    \subsection{Training Data Design}
	\label{subsec:inputDataGeneration}
    The training samples are the input for the CNN. In order to generate the training data, we select a set of target DoA locations and analyze all possible array configurations. We then generate class labels for those arrays that minimize a lower-bound on the DoA estimation error. We choose CRB to label the training samples because this bound leads to simplified expressions. {While other bounds such as BZB and WWB can also be considered \cite{antennaSelectionCognitive, antennaSelectionCognitive2,machineLearningAntennaSelection,ben2009lower}, existing literature provides their expressions only for uniform linear arrays (ULA). Hence we use CRB for UCA and RDA geometries in this work.}
    
	Consider $L$ statistically independent observations of the $q$th subarray with $K$ elements 
	\begin{align}
    \label{eq:rxq}
	\textbf{y}_q(l) = \textbf{a}_{q}(\Theta)s(l) + \textbf{n}_q(l),
	\end{align}
	where $\textbf{a}_{q}(\Theta)$ and $\textbf{n}_q(l)$ denote the $K\times 1$ elements of $\textbf{a}(\Theta)$ and $\textbf{n}(l)$ corresponding to the $q$th subarray position set $\mathcal{S}_q$. The signal and noise are assumed to be stationary and ergodic over the observation period. The covariance matrix of the observations for the $q$th subarray is
	\begin{align}
    \label{eq:analytCovar}
	\textbf{R}_q &= \text{E}\left\lbrace \textbf{y}_q\textbf{y}_q^H \right\rbrace = \textbf{a}_{q}(\Theta)E\{s[l]s^H[l]\}\textbf{a}_{q}^H(\Theta) + \sigma^2_n\textbf{I}_K,
	\end{align}
where $\textbf{I}_K$ is the identity matrix of dimension $K$. In order to simplify the CRB expressions, we represent the $K\times 1$ steering vector $\textbf{a}_q(\Theta)$ as $\textbf{a}_q$, and assume that $E\{s[l]s^H[l]\} = \sigma^2_s$ where $\sigma
^2_s$ and $\sigma^2_n$ are known. {Let $\sigma_s^2=1$ for simplicity and define SNR as $10\log_{10}(\frac{\sigma_s^2}{\sigma_n^2})$ dB.}
For this model, the CRBs for jointly estimating the target DoA coordinates $\theta$ and $\phi$ are, respectively, \cite{crbStoicaNehorai,friedlander,ye2008two}\par\noindent\small
	\begin{align}
	\textbf{CRB}_{\theta} = \frac{\sigma_n^2}{2L\operatorname{\mathbb{R}e}\{ (\dot{\textbf{a}}_{q_{\theta}}^H [\textbf{I}_K - \textbf{a}_q \textbf{a}_q^H/K ]\dot{\textbf{a}}_{q_{\phi}}) \odot (\sigma_s^4\textbf{a}_q^H\textbf{R}_q^{-1} \textbf{a}_q)\}},
    \end{align}\normalsize
   and\par\noindent\small
   \begin{align}
   \textbf{CRB}_{\phi} &= \frac{\sigma_n^2}{2L\operatorname{\mathbb{R}e}\{ (\dot{\textbf{a}}_{q_{\phi}}^H [\textbf{I}_K - \textbf{a}_q \textbf{a}_q^H/K ]\dot{\textbf{a}}_{q_{\theta}}) \odot (\sigma_s^4\textbf{a}_q^H\textbf{R}_q^{-1} \textbf{a}_q)\}}. 
    \end{align}\normalsize
The partial derivatives $\dot{\textbf{a}}_{q_{\theta}} = \frac{\partial \textbf{a}_q}{\partial_{\theta}}$ and $\dot{\textbf{a}}_{q_{\phi}} = \frac{\partial \textbf{a}_q}{\partial_{\phi}}$ are computed using the expressions in (\ref{eq:stevec1})-(\ref{eq:stevec2}). The absolute CRB \cite{ye2008two} for the two-dimensional DoA $\Theta = \{\theta,\phi\}$ using subarray $\mathcal{S}_q$ is \par\noindent\small
	\begin{align}
    \label{eq:abscrb}
	\eta(\Theta,\mathcal{S}_q) = \frac{1}{\sqrt{2}}(\textbf{CRB}_{\theta}^2 + \textbf{CRB}_{\phi}^2)^{1/2}.
	\end{align}\normalsize

	\begin{table}[H]
		\processtable{Number of classes $Q$ and the reduced number of classes $\bar{Q}$ for the uniform circular array (UCA) geometry with $M = 10$ and $M=16$ antennas. Here, elevation angle is fixed at $\theta=90^{\circ}$ and the number of azimuthal grid points $P_{\phi}=100$ for uniformly gridded azimuth plane in $[0^{\circ}, 360^{\circ})$. \label{tableComparisonForNumberOfClasses}}
		{\begin{tabular}{ccccccc}
				\hline
				
				&$K=3$ &$K=4$ &$K=5$ &$K=6$ &$K=7$ &$K=8$  \\
				\hline
				\multicolumn{7}{c}{UCA with $M=10$} \\
				\footnotesize $Q$&$120$ &$210$ &$252$ &$210$ &$120$ &$45$  \\
				\footnotesize $\bar{Q}$&$10$ &$10$ &$10$ &$10$ &$10$ &$9$  \\
				\hline
				\multicolumn{7}{c}{UCA with $M=16$} \\
				
				\footnotesize	$Q$& $560$ &$1820$ &$4368$ &$8008$ &$11440$ &$12870$  \\
				\footnotesize	$\bar{Q}$&$16$ &$10$ &$16$ &$11$ &$16$ &$16$  \\
				\hline
				
				\hline
		\end{tabular}}{}
	\end{table}\vspace{-12pt}
	{The classification problem for antenna selection poses a large number of classes especially for large arrays since $Q$ increases  significantly with $O(M!)$. To alleviate this issue, one can collect the classes randomly to reduce the complexity according to a computation/performance trade-off \cite{bibal2017ClassAmbiguity}. Due to the direction of the target and the array geometry, $\bar{Q}$, the number of classes that provides best subarrays, is much smaller than $Q$ which allows us to reduce the number of classes.}
    
    In order to label the training samples, {we first compute the sample covariance matrix from $L$ snapshots of noisy observations. We then obtain} the CRB $\eta (\Theta,{\mathcal{S}_q})$ for each target direction $\Theta$ in the training set with all subarrays $q=1,\dots, Q$. The class labels for the input data indicate the \textit{best array}, i.e. the array which minimizes the CRB in a given scenario. Let us define $\bar{Q}$ as the number of subarrays that provide the best DoA estimation performance for different directions. Then, $\bar{Q}$ is generally much smaller than $Q$ because of the direction of the target and the aperture of the subarrays. For an illustrative comparison of $Q$ and $\bar{Q}$, we refer the reader to Table \ref{tableComparisonForNumberOfClasses} which lists the number of these classes for a uniform circular array. Hence, we construct a new set $\mathcal{L}$ which includes only those classes that represent the selected subarrays for different directions\par\noindent\small
	\begin{align}
	\mathcal{L} = \{ l_1,l_2,\dots, l_{\bar{Q}}  \},
	\end{align}\normalsize
	where $\bar{Q}$ is the reduced number of classes: $l_{\bar{q}}$ is the subarray class that provides the lowest CRB for direction $\Theta$, namely\par\noindent\small
	\begin{align}
	\label{computeCRB}
	l_{\bar{q}} = \arg \min_{q=1,\dots,Q} \eta(\Theta,\mathcal{S}_q),
	\end{align}\normalsize
	for $\bar{q} = 1,\dots, \bar{Q}$.  Once the label set $\mathcal{L}$ is obtained, the input-output data pairs are constructed as $(\textbf{X}, z)$ where $\textbf{X}$ is a $M\times M\times 3$ real-valued input data obtained from the covariance matrix as defined in Sec. \ref{subsec:network} and $ z \in \mathcal{L}$ is the output label which represents the best subarray index for the covariance input $\textbf{R}$. 
    
    We summarize the steps for generating the training data in Algorithm \ref{alg:algorithm}. In step 4 of the Algorithm, the class $\mathcal{L}$ is chosen from the full combination {${Q}$}. For large size arrays, one could train by choosing less correlated subarrays or even randomly dropping out some of the subarrays. The covariance matrix used in the computation of the CRB in step 4 is the sample data covariance $\hat{\textbf{R}}_q = \frac{1}{L} \sum_{l=1}^{L} \textbf{y}_q(l)\textbf{y}_q^H(l)$ {generated with SNR$_{\textrm{TRAIN}}$}. Even though an analytical expression for $\textbf{R}_q$ is available, we use the sample data covariance here because it is closer to a practical radar operation where, in general, $\textbf{R}_q$ is estimated.
    	\begin{algorithm}[h]
		\begin{algorithmic}[1]
        	\caption{Training data generation.}
        	\Statex {\textbf{Input:} Number of given antenna elements $M$, number of desired elements $K$, number of snapshots $L$, number of different DoA angles considered $P$, number of signal and noise realizations $T$, {SNR$_{\textrm{TRAIN}}$ and $\sigma_s^2=1$.}  }
   			\label{alg:algorithm}
            \Statex {\textbf{Output:} Training data: Input-output pairs consisting of sample covariances {$\hat{\textbf{R}}^{(i,p)}$} and output labels $z_{p}^{(i)}$ for $p=1,\dots,P$ and $i = 1,\dots,T$.}
			\State Select $P$ DoA angles $\Theta_p = (\theta_p,\phi_p)$ for $p =1,\dots,P$. 
			\State Generate $T$ different realizations of the array output $\{\textbf{Y}_{p}^{(i)}\}_{i=1}^T$ for $p = 1,\dots,P$ as\par\noindent\small
			\begin{align}
			\textbf{Y}_p^{(i)} = [\textbf{y}_p^{(i)}(1),\dots,\textbf{y}_p^{(i)}(L) ], \nonumber
			\end{align}\normalsize
			where $\textbf{y}_p^{(i)}(l) = \textbf{a}(\Theta_p)s^{(i)}(l) + \textbf{n}^{(i)}(l)$, $s^{(i)}(l) \sim \mathcal{CN}(0,\sigma_s^2)$ and $\textbf{n}^{(i)}(l) \sim \mathcal{CN}(0,\sigma_n^2\textbf{I})$.
			\State { Construct the sample covariance matrix $\hat{\textbf{R}}$ and the $K\times K $ covariance matrices $\hat{\textbf{R}}_q^{(i,p)}$ for $q = 1,\dots,Q$.}
            \State Compute the CRB values $\eta(\Theta_p,\mathcal{S}_q)$ following (\ref{eq:abscrb}) and obtain the class set $\mathcal{L}$ representing the best subarrays using (\ref{computeCRB}).
			\State {Generate the input-output pairs as $(\hat{\textbf{R}}^{(i,p)}, z_{p}^{(i)})$ for $p=1,\dots,P$ and $i = 1,\dots,T$}.
            \State Construct training data by concatenating the input-output pairs:\par\noindent\small
		{\begin{align}
			\mathcal{D}_{\text{train}} = \{ ({\hat{\textbf{R}}^{(1,1)}}, z_{1}^{(1)}), ({\hat{\textbf{R}}^{(2,1)}}, z_{1}^{(2)}),\dots, ({\hat{\textbf{R}}^{(T,1)}}, z_{1}^{(T)}), \nonumber \\ ({\hat{\textbf{R}}^{(1,2)}}, z_{2}^{(1)})\dots, ({\hat{\textbf{R}}^{(T,P)}}, z_{P}^{(T)})\}, \nonumber
			\end{align}\normalsize}
			where the size of the training dataset is $J = TP$.			
		\end{algorithmic}
	\end{algorithm}
	\subsection{Network Structure and Training}
    \label{subsec:network}
    Using the labeled training dataset, we build a trained CNN classifier. The input of this learning system is the data covariance and the output is the index of the selected antenna set.
    
Given the $M \times L$ output $\textbf{Y}$ of the antenna array, the corresponding sample covariance is a complex-valued $M\times M$ matrix $\hat{\textbf{R}}$. The first step towards efficient classification is to define a set of real-valued features that capture the distinguishing aspects of the output. The features we consider in this work are the angle, real and imaginary parts of $\hat{\textbf{R}}$. One could also consider magnitude here but we did not find much difference in the results when this feature was included. We construct three $M\times M$ real-valued matrices $\{\textbf{X}_c\}_{c=1}^{3}$ whose $(i,j)$th entry contain, respectively, the phase, real and imaginary parts of the signal covariance matrix $\hat{\textbf{R}}$:
    $[\textbf{X}_{1}]_{i,j} = \angle[\hat{\textbf{R}}_{}]_{i,j}$; $[\textbf{X}_{2}]_{i,j} = \operatorname{\mathbb{R}e}\left\lbrace [\hat{\textbf{R}}_{}]_{i,j}\right\rbrace$; and $[\textbf{X}_{3}]_{i,j} = \operatorname{\mathbb{I}m}\left\lbrace [\hat{\textbf{R}}_{}]_{i,j}\right\rbrace$. 
	
\begin{figure*}[ht]
		\centering
		{\includegraphics[width=.6\textheight]{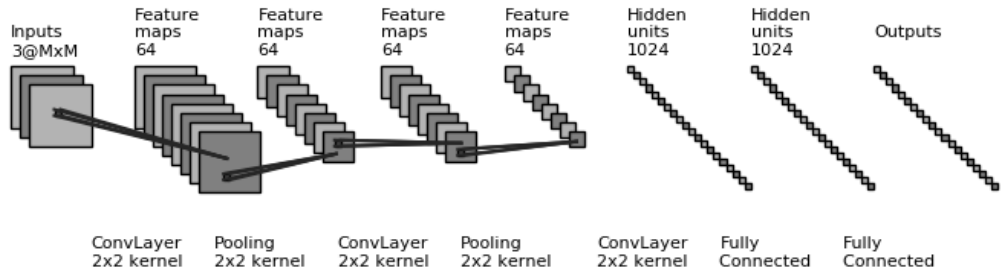}}
		\caption{CNN structure for antenna selection.}
		\label{figNetwork}
	\end{figure*}

    Figure~\ref{figNetwork} depicts the deep learning CNN structure that we used. The proposed network consists of 9 layers. In the first layer, the CNN accepts the two-dimensional inputs $\{\textbf{X}_c\}_{c=1}^{3}$ in three real-valued channels. The second, fourth and sixth layers are convolutional layers with 64 filters of size $2\times 2$. The third and fifth layers are max-pooling to reduce the dimension by 2. 
The seventh and eighth layers are fully connected with 1024 units whose $50\%$ are randomly dropped out to reduce overfitting in training \cite{srivastavaDropoutLayer}. There are rectified linear units (ReLU) after each convolutional and fully connected layers where the $\text{ReLU}(x) = \text{max}(x,0)$. At the output layer, there are $Q$ units wherein the network classifies the given input data using a softmax function and reports the probability distribution of the classes to provide the best subarray.

	In order to train the proposed CNN, we sample the target space for $P$ directions and collect the data for several realizations. We realized the proposed network in MATLAB on a PC with 768-core GPU. During the training process, $90\%$ and $10\%$ of all data generated are selected as the training and validation datasets, respectively. Validation aids in hyperparameter tuning during the training phase to avoid the network simply memorizing the training data rather than learning general features for accurate prediction with new (test) data. We used the stochastic gradient descent algorithm with momentum \cite{bishop2006pattern} for updating the network parameters with learning rate $0.05$ and mini-batch size of $500$ samples for 50 epochs. As a loss function, we use the negative log-likelihood or cross-entropy loss.
Another useful metric we consider for evaluating the network is the accuracy:\par\noindent\small
	\begin{align}
	\label{eq:accuracy}
		\text{Accuracy } (\%) = \frac{{ \zeta}}{J}\times 100,
    \end{align}\normalsize
where $J$ is the total number of input datasets in which the model identified the best subarrays correctly $\zeta$ times. This metric is available for training, validation and test phases.
	\vspace{-14pt}
	\section{Numerical Experiments}
	\label{sec:Simulations}
    In this section, we present numerical experiments to train and test the proposed CNN structure shown in Fig. \ref{figNetwork} for different antenna geometries. In the following, we append subscripts TEST and TRAIN to indicate parameter values used for training and testing modes, respectively. {The training data is obtained by sampling the DoA space with $P$ directions whereas the DoA angles in the test data are uniform randomly selected.} 

\subsection{Uniform Linear Array}
\label{subsec:ula}
We first analyze the effect of the performance metrics on the antenna selection and DoA estimation accuracy by employing the simplest and most common geometry of a ULA. For creating the training data, we employed three bounds: CRB, BZB and WWB \cite{performanceBoundsWWB}. The network was trained for $M=10$, $K=4$, $L_{\text{TRAIN}}=100$ snapshots, $T_{\text{TRAIN}}=100$ signal and noise realizations, and $P_{\text{TRAIN}}=100$ DoA angles. The number of uniformly spaced azimuthal grid points are set to $P_{\phi} = 100$.
	\begin{figure}[ht]
		\centering
		{\includegraphics[width=.25\textheight,height=.15\textheight]{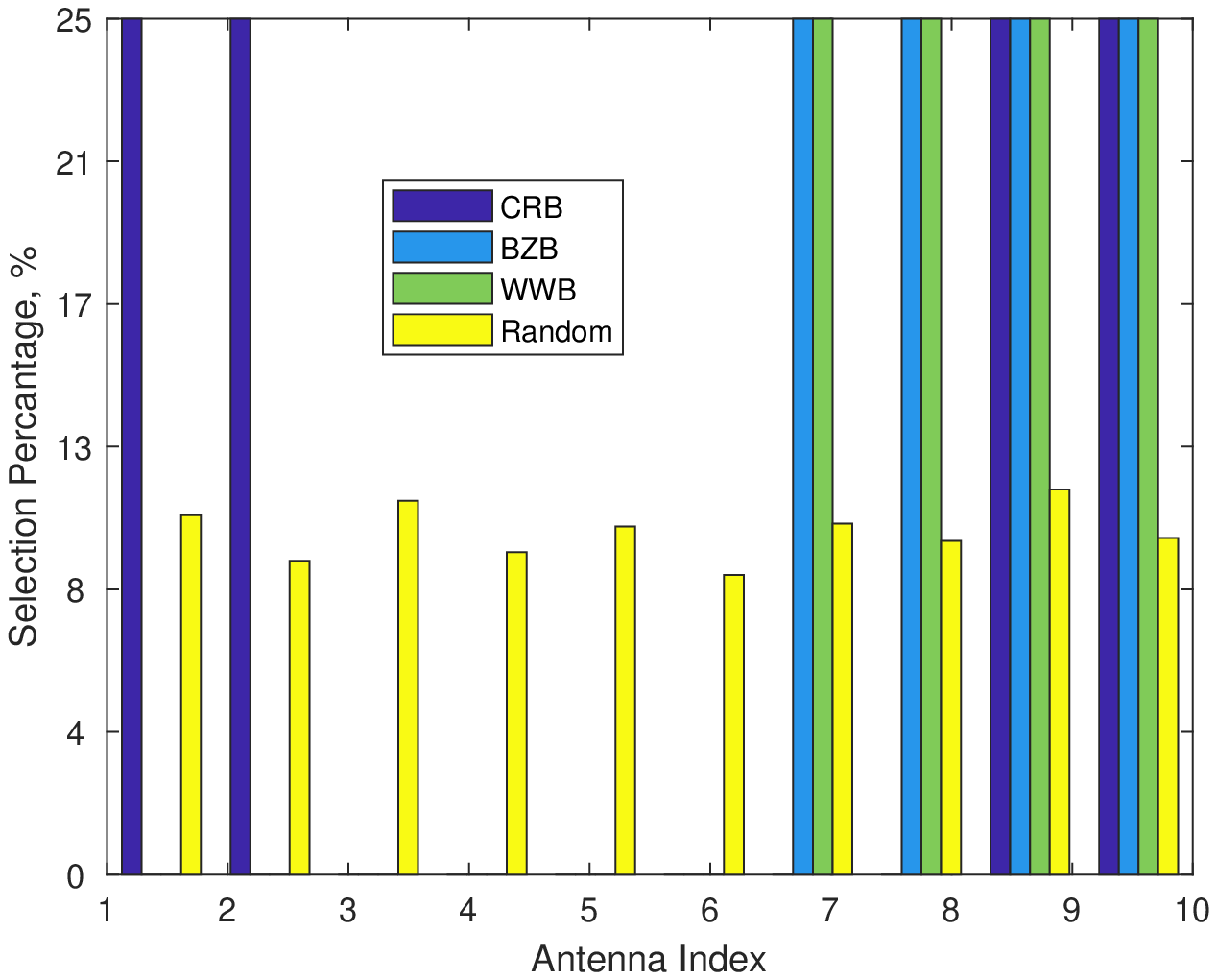}} \\
			{\includegraphics[width=.25\textheight,height=.15\textheight]{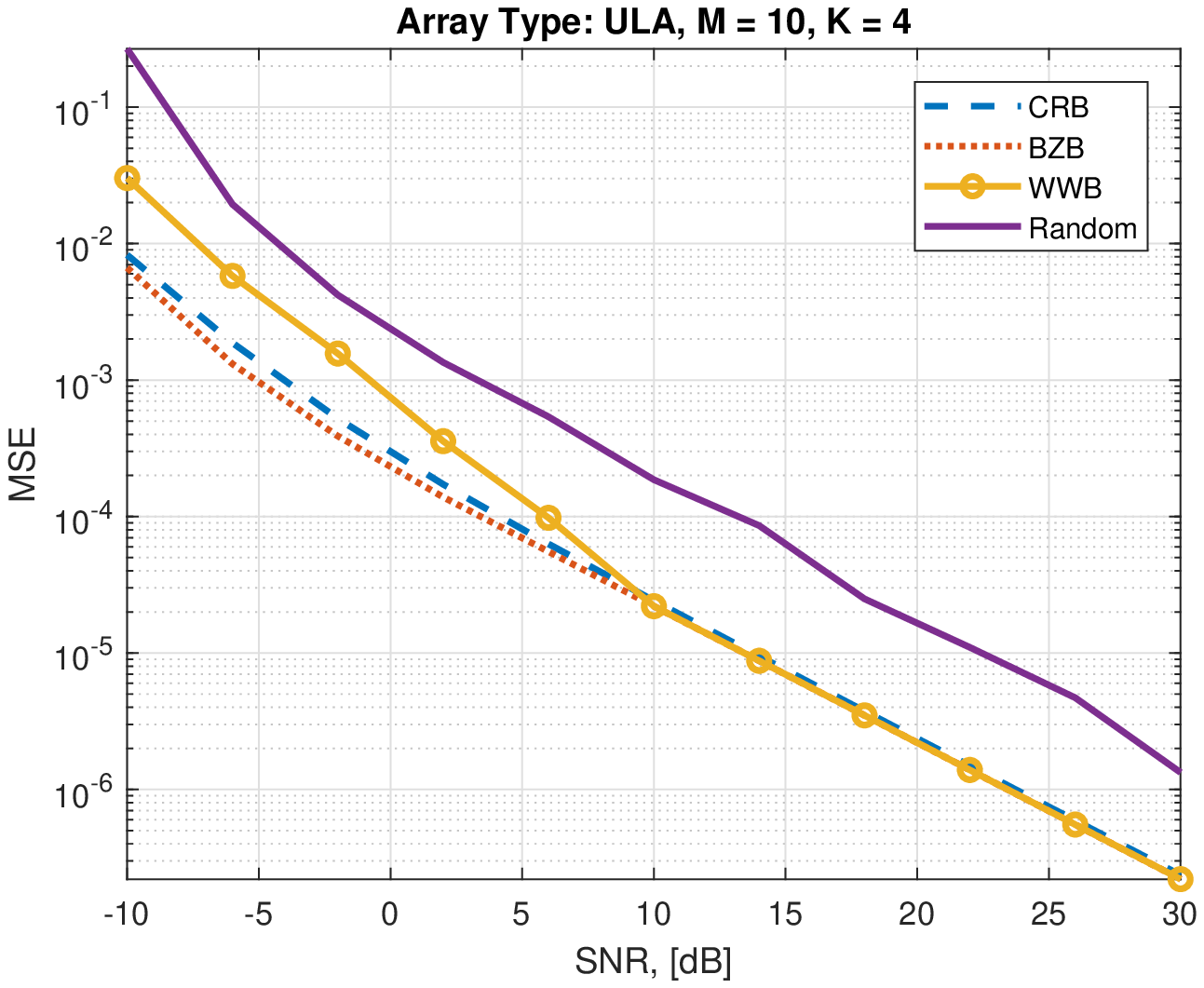}}
		\caption{Top: Antenna selection percentage over $J_{\text{TEST}}=10000$ trials. Bottom: MSE of DoA for selected subarrays. The array geometry is a ULA with $M=10$ and $K=4$.}
		\label{figULABoundAnalysis}
	\end{figure}

For test mode, we fed the network with data corresponding to $P_{\text{TEST}}=100$ DoA angles different than the ones used in the training phase but keeping the values of $M$, $K$, $L$ and $T$ same as in the training. The top plot of Fig.~\ref{figULABoundAnalysis} shows the percentage of times a particular antenna index appears as part of the optimal array in the output over $J_{\text{TEST}}=T_{\text{TEST}}P_{\text{TEST}}$ trials with different performance metrics used during training. As seen here, when the CNN is trained with data created from the CRB, the classifier output arrays usually consists of the elements at the extremities. However, the network trained on BZB and WWB usually selects arrays with elements close to each other leading to low sidelobe levels. Also shown here is the random selection wherein each element is chosen with approximately $10\%$ selection rate. We provide the DoA estimation performance of the antenna subarrays selected by the network for different values of test data SNRs in the bottom plot of Fig.~\ref{figULABoundAnalysis}. We observe that, compared to our DL approach, the random thinning results in inferior DoA estimation due to small array aperture. Among various bounds, the MSE is somewhat similar at high SNR regimes with the BZB faring better than CRB at low SNRs. 

\subsection{2-D Arrays}
\label{subsec:2da}
    We now investigate more complicated array geometries such as uniform circular arrays (UCAs) and randomly deployed arrays (RDAs). In Table \ref{tableComparisonForNumberOfClasses}, the computed values of $Q$ and the reduced number of classes $\bar{Q}$ are shown for UCAs with $M=10$ and $M=16$ elements where {$P_{\phi}=100$} are uniformly spaced grid points in $[0^{\circ}, 360^{\circ})$ and $\theta=90^{\circ}$. We remark that the size of best subarray classes, $\bar{Q}$, is much less than $Q$. 

	\begin{figure*}[ht]
		\centering
		\subfloat[]{\includegraphics[width=.24\textheight,height=.17\textheight]{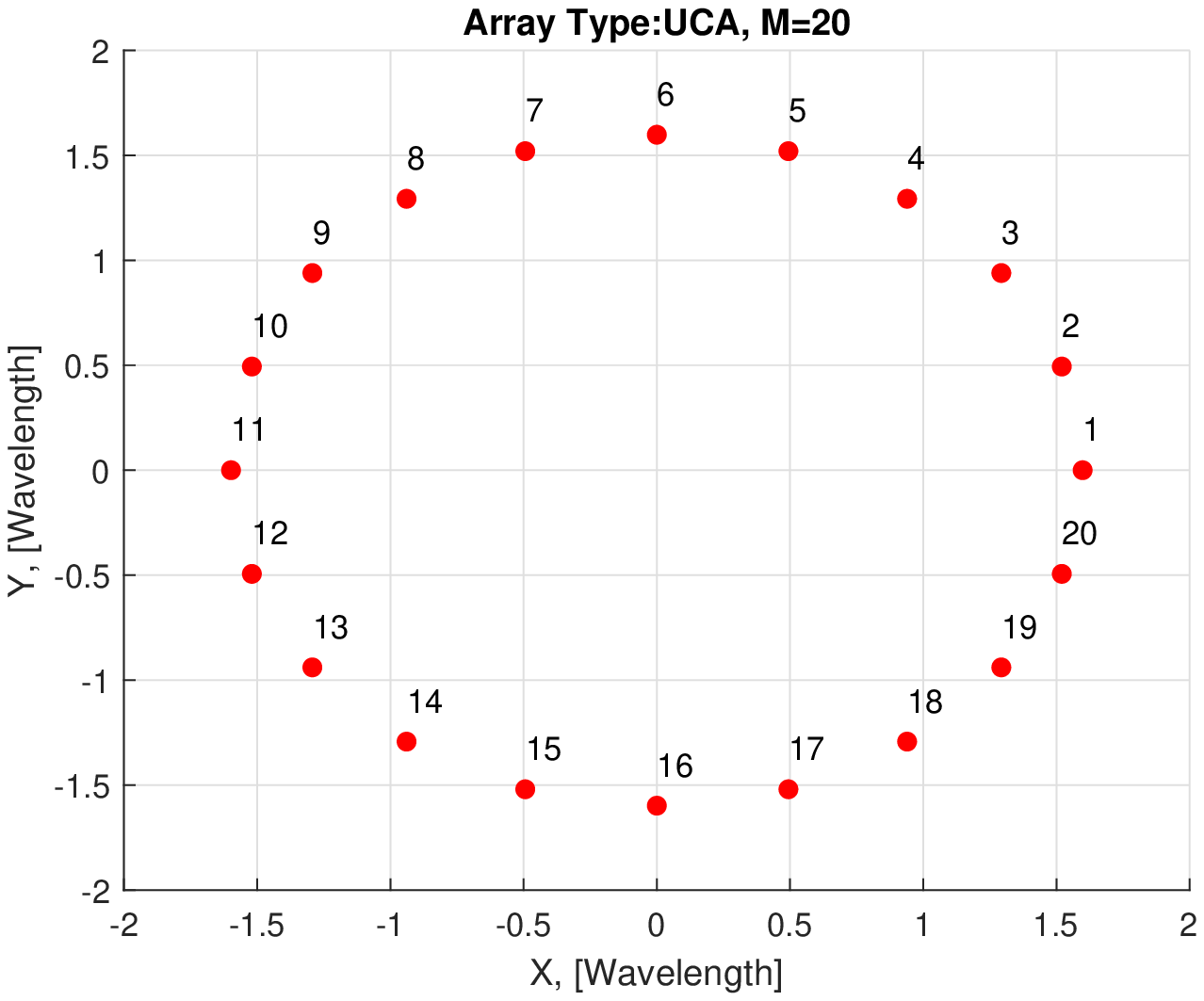}%
			\label{figPosUCAM20} }
		\subfloat[]{\includegraphics[width=.24\textheight,height=.17\textheight]{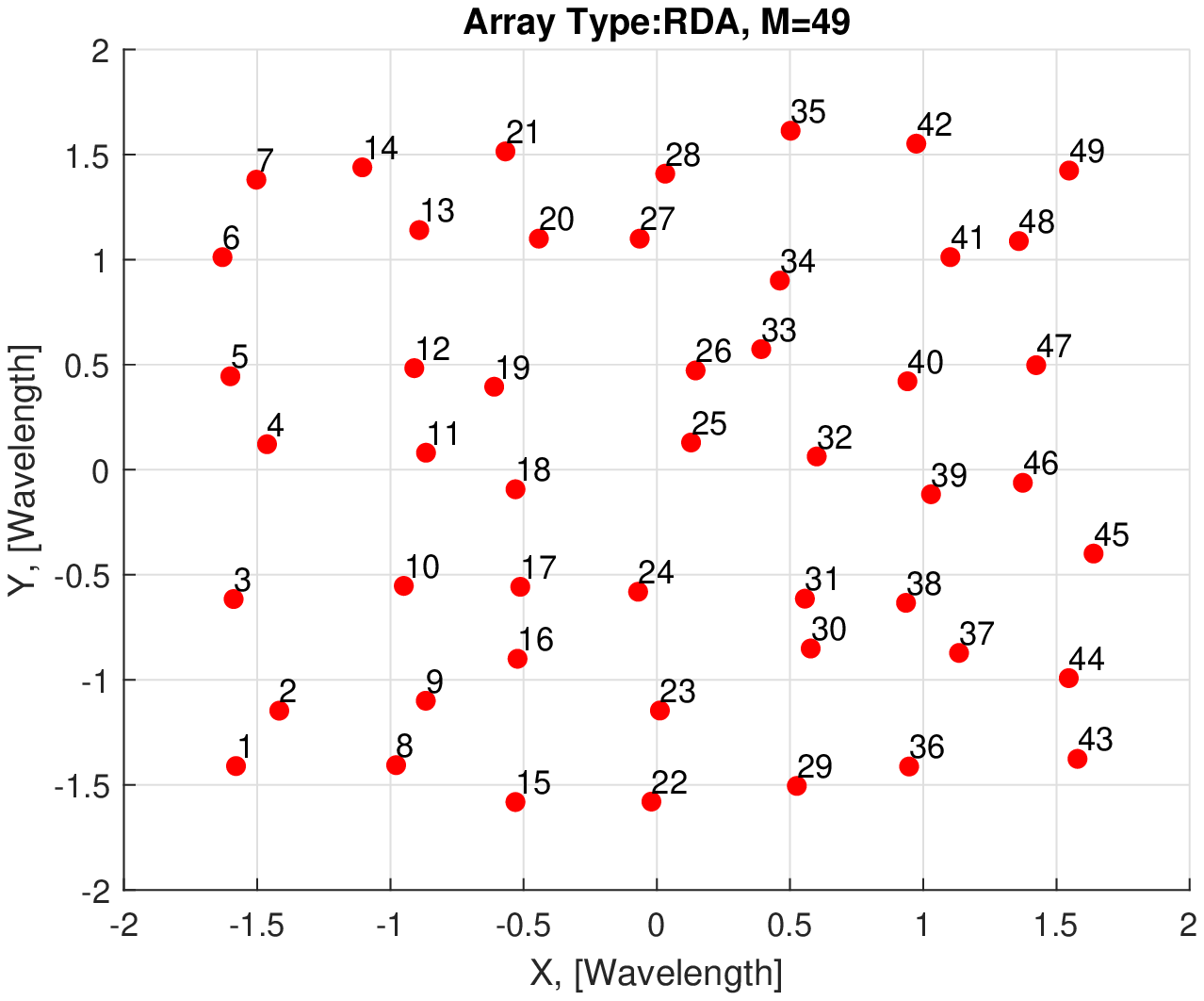}%
			\label{figPosRAM49} }
		\subfloat[]{\includegraphics[width=.24\textheight,height=.17\textheight]{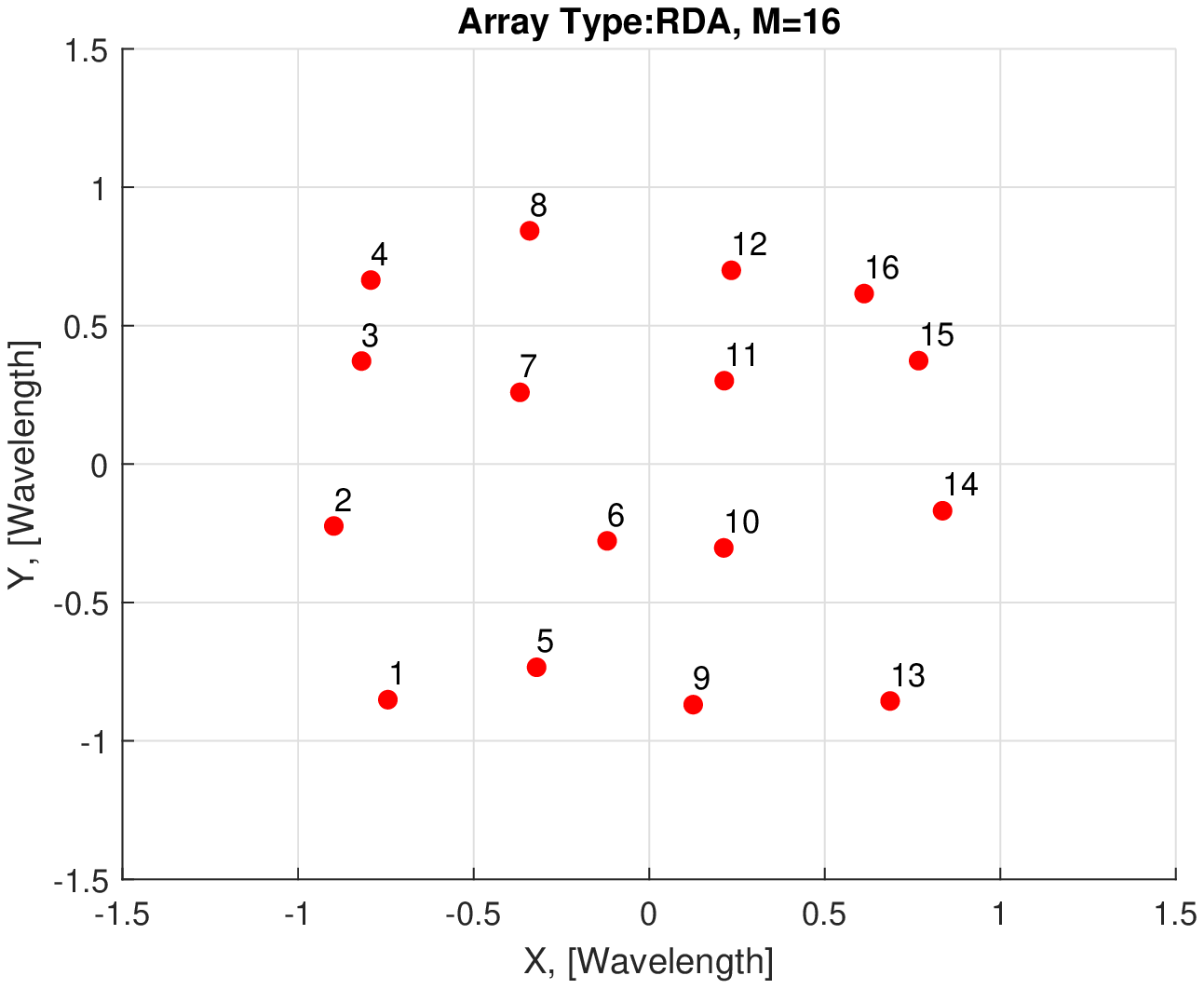}%
			\label{figPosRAM16} }
		\caption{Placement of antennas for a) UCA with $M=20$ elements, b) RDA with $M=49$ and c) RDA with $M=16$ .}
		\label{figArrayPlacement1D}
	\end{figure*}
	
	\subsubsection{Experiment \#1: 1-D Scenario}
	\begin{table*}[t]
		\processtable{Selected antenna indices for UCA with $M = 16$ and $K \in \{3,4,5,6,7\}$, $\phi \in \{21.42^{\circ},74.28 ^{\circ},127.14^{\circ},232.85 ^{\circ},285.71 ^{\circ}\}$. The number of snapshots $L = 100$ and SNR$_{\text{TRAIN}}$ =20dB. The antennas are indexed counter-clockwise with the first antenna placed as shown in Fig. \ref{figPosUCAM20}. \label{tableSelectedAntennaIndexes1D}}
		{\begin{tabular}{ccccccc}
			& $21.42^{\circ}$ 
			
			&  $74.28 ^{\circ}$
			
			& $127.14 ^{\circ}$
			
			& $ 232.85 ^{\circ}$
			
			&  $285.71 ^{\circ}$
			
			\\
			
			\hline
			K=3
			&\footnotesize $\{ 7 ,  17  ,  18\}$   
			&\footnotesize $\{10 ,   11 ,   20\}$   
			&\footnotesize $\{3  ,  13  ,  14\}$
			&\footnotesize $\{9  ,  18  ,  19\}$  
			&\footnotesize $\{2  ,  11,    12\}$ 
			\\
			K=4
			&\footnotesize $\{7 ,    8  ,  17 ,   18\}$   
			&\footnotesize $\{1   , 10 ,   11   , 20\}$   
			&\footnotesize $\{3   ,  4,    13   , 14\}$
			&\footnotesize $\{8   ,  9,    18   , 19\}$  
			&\footnotesize $\{1   ,  2,    11   , 12\}$ 
			\\
			K=5
			&\footnotesize $\{ 7  ,   8 ,   16,    17,    18\}$   
			&\footnotesize $\{1   , 10 ,   11 ,   19,    20\}$   &\footnotesize $\{3   ,  4 ,   12 ,   13,    14\}$
			&\footnotesize $\{8   ,  9 ,   10 ,   18,    19\}$  
			&\footnotesize $\{1   ,  2 ,   11 ,   12,    13\}$ 
			\\
			K=6
			&\footnotesize $\{6 ,    7 ,    8  ,  16 ,   17  ,  18\}$   
			&\footnotesize $\{ 1  ,   9 ,   10  ,  11 ,   19  ,  20\}$   
			&\footnotesize $\{ 2  ,   3 ,    4  ,  12 ,   13  ,  14\}$
			&\footnotesize $\{ 8  ,   9 ,   10  ,  18 ,   19  ,  20\}$  
			&\footnotesize $\{1  ,   2 ,    3  ,  11 ,   12  ,  13\}$ 
			\\
			K=7&\footnotesize $\{ 6 ,    7,     8,    16,    17,    18,    19\}$   
			&\footnotesize $\{1  ,   9 ,   10 ,   11 ,   12 ,   19 ,   20\}$   
			&\footnotesize $\{2  ,   3 ,    4 ,    5 ,   12  ,  13 ,   14\}$
			&\footnotesize $\{8 ,    9 ,   10 ,   17 ,   18  ,  19 ,   20\}$  
			& \footnotesize $\{1 ,    2 ,    3 ,   11 ,   12  ,  13  ,  20
			\}$ 
			\\
			\hline
		\end{tabular}}{}
	\end{table*}
	\begin{table*}[t]
		\processtable{The accuracy percentages for training and validation datasets in 1-D and 2-D scenario. \label{tableAccuracies}}
		{\begin{tabular}{c|cc|cc|cc|cc}
			\hline
			&\multicolumn{2}{|c|}{1-D, UCA with $M=20$, $K=6$.} &\multicolumn{2}{c}{1-D, UCA with $M=45$, $K=5$.} & \multicolumn{2}{|c}{1-D, RDA with $M=49$, $K=5$.} &\multicolumn{2}{|c}{2-D, RDA with $M=16$, $K=6$.}\\
			\hline
			SNR$_{\text{TRAIN}}$ & Training & Validation & Training  & Validation  & Training  & Validation & Training  & Validation \\
			10 dB &65.2\%  & 68.7\% & 98.7\% &97.8\%  &97.8\% & 95.7\% &8.1\% & 10.7\%\\
			15 dB &98.1\%  & 98.5\% & 99.0\% &98.7\%   &99.9\%  &  98.6\%&60.1\%  &  63.2\%\\
			20 dB & 99.2\%& 99.5\%  & 100\% &99.7\%  & 97.5\% & 98.1\% &80.8\% & 80.6\%\\
			25 dB & 99.4\% & 99.8\%  & 100\% &100\%   & 100\% & 100\% &88.9\% & 89.2\%\\
			30 dB & 100\% & 100\%  &100\%  &100\%   & 100\% & 100\% & 82.6\% & 83.2\% \\
			inf dB & 100\% & 100\%  & 100\% &100\%   &100\%  & 100\% &85.0\%  & 83.9\%\\
			\hline
		\end{tabular}}
		
	\end{table*}
    \begin{table*}[t]
		\caption{Selected antenna indices for random array with $M = 16$ and $K \in \{3,4,5,6,7\}$, $\phi \in \{30^{\circ},60 ^{\circ},90^{\circ},120 ^{\circ}, 210 ^{\circ}\}$ and $\theta \in \{ 90^{\circ},92^{\circ} \}$. The number of snapshots $L = 100$ and SNR$_{\text{TRAIN}}$=20dB. The antenna indices are given in Fig. \ref{figPosRAM16}.}
		\label{tableSelectedAntennaIndexes2DScenario}
		\centering
		
		\begin{tabular}{|c|c|c|c|c|c|c|}
			\hline
			\hline
			&\footnotesize $\theta = 90^{\circ},\phi =30^{\circ}$ 
			
			& \footnotesize $\theta = 90^{\circ},\phi =60 ^{\circ}$
			
			&\footnotesize $\theta = 90^{\circ},\phi =90 ^{\circ}$
			
			&\footnotesize $\theta = 90^{\circ},\phi = 120^{\circ}$
			
			& \footnotesize $\theta = 90^{\circ},\phi =210 ^{\circ}$
			
			\\
			
			\hline
			K=3
			&\footnotesize $\{4   ,  8 ,   13\}$   
			&\footnotesize $\{3   ,  4 ,   13\}$   
			&\footnotesize $\{2   , 14 ,   15\}$
			&\footnotesize $\{1   , 15 ,   16\}$  
			&\footnotesize $\{4   ,  8 ,   13\}$ 
			\\
			\hline 
			K=4
			&\footnotesize $\{4  ,   8  ,   9   , 13\}$   
			&\footnotesize $\{3  ,   4  ,  13   , 14\}$   
			&\footnotesize $\{2  ,   3  ,  14   , 15\}$
			&\footnotesize $\{1  ,   2  ,  15   , 16\}$  
			&\footnotesize $\{4  ,   8  ,   9   , 13\}$ 
			\\
			\hline 
			K=5
			&\footnotesize $\{3,     4,     8 ,    9 ,   13\}$   
			&\footnotesize $\{3,     4,     8 ,   13 ,   14\}$  
			&\footnotesize $\{2,     3,     4 ,   14 ,   15\}$
			&\footnotesize $\{1,     2,    14 ,   15 ,   16\}$  
			&\footnotesize $\{2,     3,     4 ,   14 ,   15\}$
			\\
			\hline 
			K=6
			&\footnotesize $\{3   ,  4  ,   8  ,   9  ,  13  ,  14\}$   
			&\footnotesize $\{ 3  ,   4 ,    8 ,    9 ,   13 ,   14\}$   
			&\footnotesize $\{ 2  ,   3 ,    4 ,   13 ,   14 ,   15\}$
			&\footnotesize $\{1   ,  2  ,   5  ,  14  ,  15  ,  16\}$  
			&\footnotesize $\{3   ,  4  ,   8  ,   9  ,  13  ,  14\}$ 
			\\
			\hline
			K=7
			&\footnotesize $\{3,     4 ,    5 ,    8  ,   9 ,   13 ,   14\}$   
			&\footnotesize $\{2,     3 ,    4 ,    8  ,   9 ,   13 ,   14\}$   
			&\footnotesize $\{1,     2 ,    3 ,    4  ,  13 ,   14  ,  15\}$
			&\footnotesize $\{1,     2 ,    5 ,   12  ,  14 ,   15 ,   16\}$  
			& \footnotesize $\{1,     2,     3,     4 ,   13,    14,    15\}$ 
			\\
			\hline
		\end{tabular}
		\begin{tabular}{|c|c|c|c|c|c|c|}
			\hline
			&\footnotesize $\theta = 92^{\circ},\phi =30^{\circ}$ 
			
			& \footnotesize $\theta = 92^{\circ},\phi =60^{\circ}$
			
			&\footnotesize $\theta = 92^{\circ},\phi =90^{\circ}$
			
			&\footnotesize $\theta = 92^{\circ},\phi = 120 ^{\circ}$
			
			& \footnotesize $\theta = 92^{\circ},\phi =210^{\circ}$
			
			\\
			
			\hline
			K=3
			&\footnotesize $\{1  ,   2 ,   15\}$   
			&\footnotesize $\{1  ,   8 ,   16\}$   
			&\footnotesize $\{1  ,   8 ,    9\}$
			&\footnotesize $\{4  ,  12 ,   13\}$  
			&\footnotesize $\{1  ,   2 ,   15\}$ 
			\\
			\hline 
			K=4
			&\footnotesize $\{1    , 2 ,   14,    16\}$   
			&\footnotesize $\{1    , 9 ,   12,    16\}$   
			&\footnotesize $\{1    , 8 ,   12,    13\}$
			&\footnotesize $\{4    , 8 ,    9,    13\}$  
			&\footnotesize $\{1    , 2 ,   14,    16\}$ 
			\\
			\hline 
			K=5
			&\footnotesize $\{1  ,   2 ,   14 ,   15  ,  16\}$   
			&\footnotesize $\{1  ,   5 ,    8 ,   12  ,  16\}$  
			&\footnotesize $\{1  ,   8 ,    9 ,   12  ,  13\}$
			&\footnotesize $\{4  ,   8 ,    9 ,   12  ,  13\}$  
			&\footnotesize $\{ 1 ,    2,    14 ,   15 ,   16\}$
			\\
			\hline 
			K=6
			&\footnotesize $\{  1 ,    2    , 5 ,   14  ,  15 ,   16 \}$   
			&\footnotesize $\{ 1  ,   5   ,  9 ,   12   , 15 ,   16\}$   
			&\footnotesize $\{1   ,  8   ,  9  ,  12   , 13  ,  16\}$
			&\footnotesize $\{3  ,   4 ,    5     ,8    , 9  ,  13\}$  
			&\footnotesize $\{ 1   ,  8  ,   9 ,   12,    13   , 16\}$ 
			\\
			\hline
			K=7
			&\footnotesize $\{1  ,   2  ,   3   ,  5  ,  14 ,   15 ,   16\}$   
			&\footnotesize $\{1  ,   5  ,   8   ,  9  ,  12 ,   15 ,   16\}$   
			&\footnotesize $\{1  ,   4  ,   8   ,  9  ,  12 ,   13 ,   16\}$
			&\footnotesize $\{3  ,   4  ,   5   ,  8  ,   9 ,   12 ,   13\}$  
			& \footnotesize $\{1  ,   2 ,    3  ,   5 ,   14,    15,    16\}$ 
			\\
			\hline
		\end{tabular}
	\end{table*}
	We assume that the target and the antenna array are placed in the same plane (i.e., $\theta = 90^{\circ}$). We consider different array geometries such as UCAs and RDAs as shown in Figs. \ref{figPosUCAM20} and \ref{figPosRAM49}, respectively. The UCAs consist of $M=20$ and $M=45$ elements where each antenna is placed a half wavelength apart from each other. In order to generate the RDA geometry, we first take a uniform rectangular array of size $7\times 7$, and then perturb the antenna positions as $\{p_{x_m} + \delta_x,p_{y_m} + \delta_y\}_{m=1}^{M}$ where $\delta_x, \delta_y \sim \text{u}\{[-0.1\lambda,0.1\lambda]\}$.
    
	The training set is constructed with $P_{\text{TRAIN}}=100$ DoA angles. {Note that $P_{\text{TRAIN}}=100$ is sufficient to train the network for antenna selection in the whole azimuth space}.  As an illustration of the training set generated, Table \ref{tableSelectedAntennaIndexes1D} lists the indices of a few UCA antennas that yield the best CRB for $M=16$ and different target directions as $K$ varies. We note that the subarrays that yield the largest aperture for the given target direction are selected due to the structure of the UCA. {Moreover, notice that the same antenna subarrays are selected for the symmetric angles due to the symmetric geometry}. The training samples are prepared for different SNR values (i.e., SNR$_{\text{TRAIN}}$) and the accuracy of training and validation phases is shown in Table \ref{tableAccuracies}.
	
\begin{figure}[t]
		\centering
		{\includegraphics[width=.35\textheight]{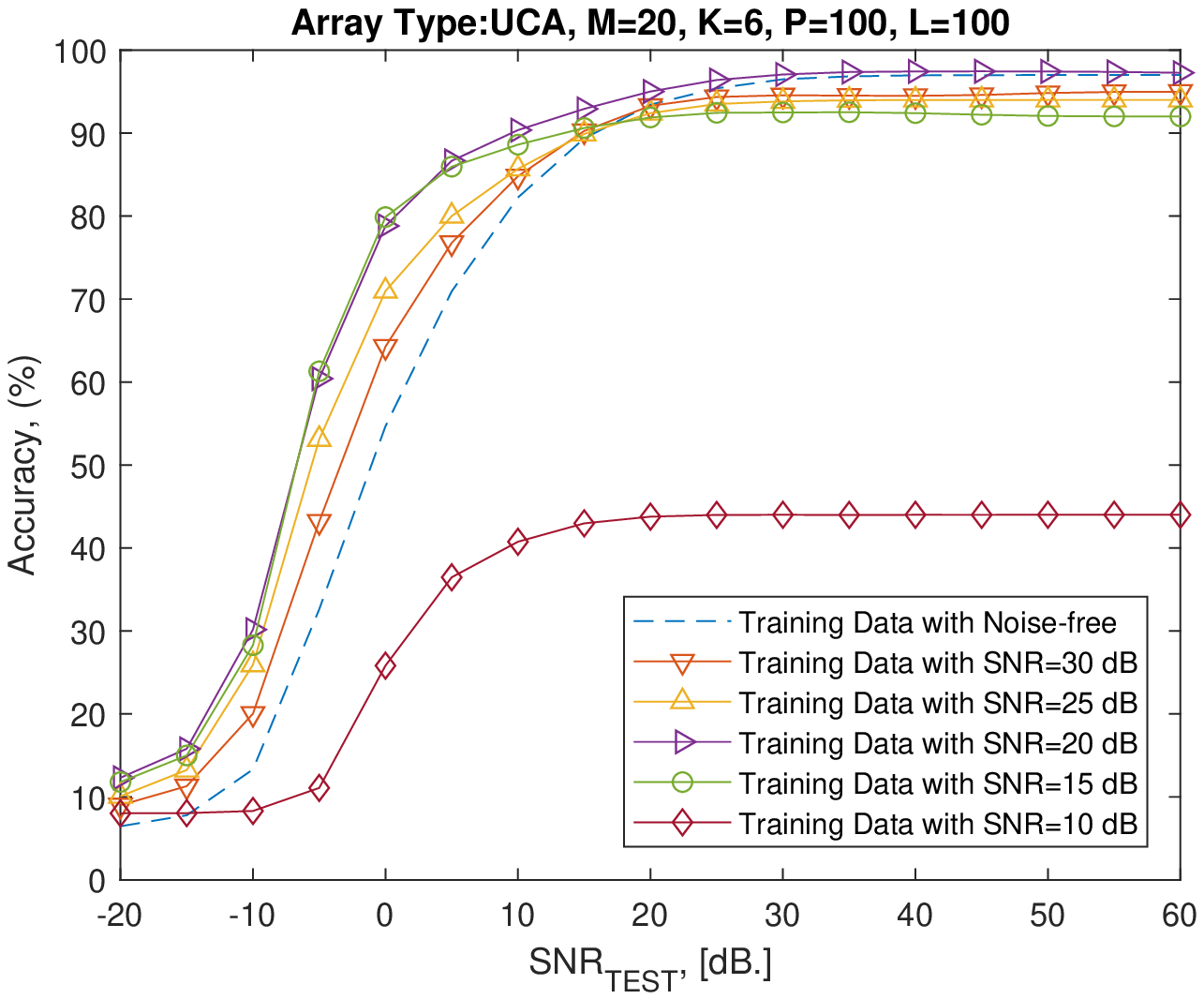}}\\
        {\includegraphics[width=.35\textheight]{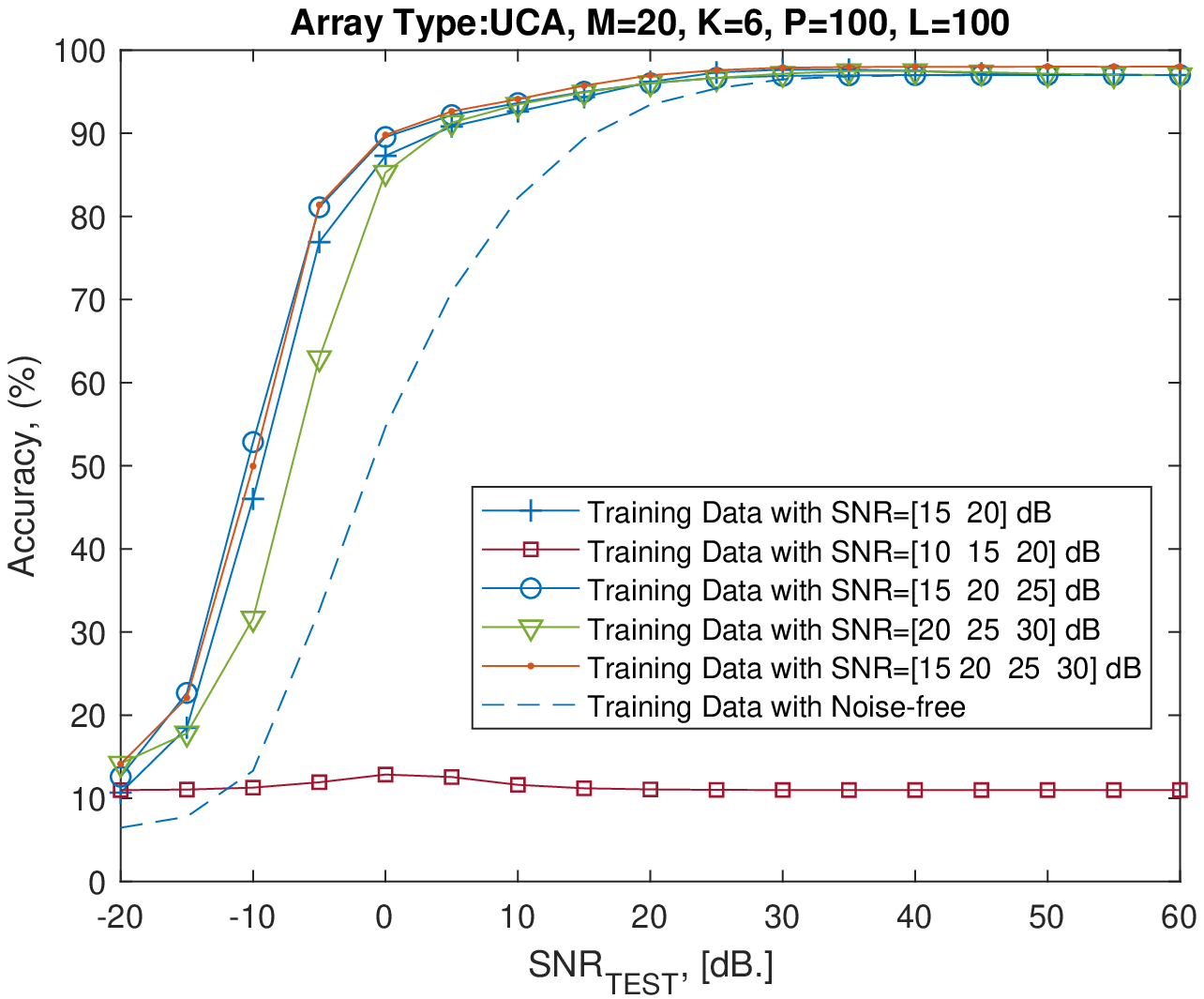}}
		\caption{{Performance of test dataset using CNN with respect to SNR$_{\text{TEST}}$. The antenna geometry is a UCA with $M=20$ and $K=6$. Top shows the performance when a single SNR value was used during the training phase. Bottom plot shows the same when multiple SNR values are used for training the network.}}
		\label{figSNRTest1DUCAM20}
	\end{figure}

	In order to evaluate the classification performance of the proposed CNN structure, we fed the trained network with the test data generated with $L_{\text{TEST}}=100$, $T_{\text{TEST}}=100$ and $P_{\text{TEST}}=100$ with $\phi_{\text{TEST}} \sim \text{u}\{[0^{\circ}, 360^{\circ}]\}$. Figure~\ref{figSNRTest1DUCAM20} shows the classification performance of the CNN for $J_{\text{TEST}}=100$ Monte Carlo trials. {The results are given for both the training generated with a single SNR$_{\text{TRAIN}}$ (top) and the multiple SNR$_{\text{TRAIN}}$ (bottom).} Figure~\ref{figSNRTest1DUCAM20} also shows the performance of the noisy test data when the network is trained with noise-free dataset; it's performance degrades especially at low SNR levels. These observations imply that noisy training datasets should be used for robust classification performance with the test data. On the other hand, when the training data is corrupted with strong noise content (e.g., SNR$_{\text{TRAIN}}\leq 10$dB), then despite using the noisy training data, the proposed CNN does not recover from poor performance at low SNR$_{\text{TEST}}$ regimes. {Similar observations can be made for multiple SNR$_{\text{TRAIN}}$ scenario, i.e., the network has poor performance if the training data includes the data prepared with SNR$_{\text{TRAIN}}=10$dB. This leads to the conclusion that the training data should not include too much noise. While there is a slight difference comparing the single and multiple SNR$_{\text{TRAIN}}$ cases for high SNR$_{\text{TEST}}$ regimes, CNN performs better in low SNR (i.e., SNR$_{\text{TEST}}\leq 10$dB ) if multiple SNRs are used in the training data. Specifically, the training data generated with SNR$_{\text{TRAIN}}$=[15 20 25 30] dB provides the best result for a large range of SNR$_{\textrm{TEST}}$.}
    
\begin{figure}[t]
		\centering
		{\includegraphics[width=.35\textheight]{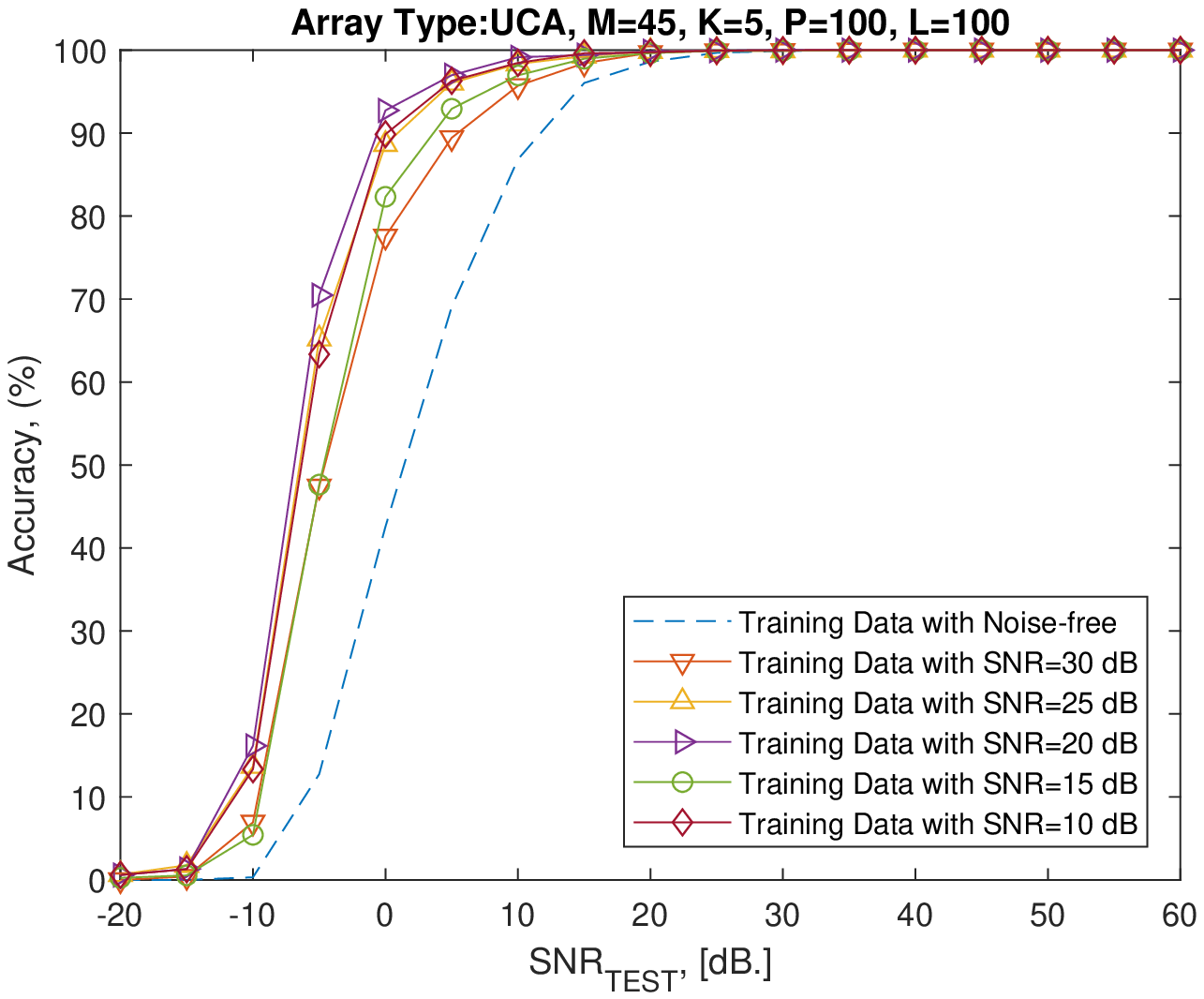}}
		\caption{Performance of test data using CNN with respect to SNR when the training data is prepared with SNR$_{\text{TRAIN}}$$\in \{10, 15, 20,25,30, \inf\}$ dB. The antenna geometry is a UCA with $M=45$ and $K=5$.}
		\label{figSNRTest1DUCAM45}
	\end{figure}

    The performance at low SNRs can be improved when the size of the array increases and, as a result, the input data is huge and the SNR is enhanced due to large $M$. As an example, Fig. \ref{figSNRTest1DUCAM45} illustrates the performance of the network for UCA with $M=45$ and $K=5$, where the network provides high accuracy for a wide range of SNR$_{\text{TEST}}$ compared to the scenario in Fig. \ref{figSNRTest1DUCAM20}. 
    
\begin{figure}[t]
	\centering
	{\includegraphics[width=.35\textheight]{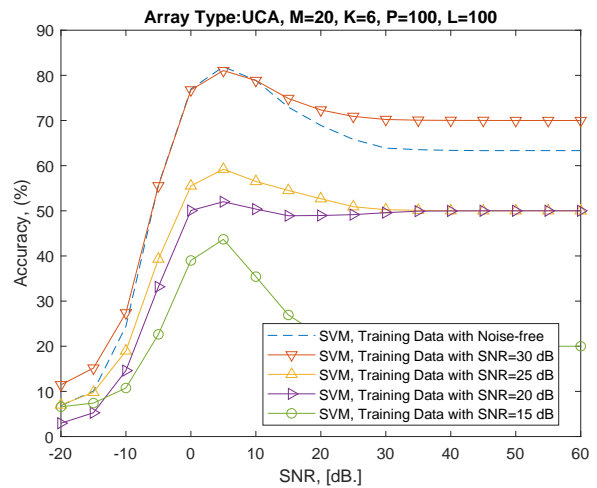}}
	\caption{Performance of test dataset using SVM with respect to SNR$_{\text{TEST}}$ when the training data is prepared with SNR$_{\text{TRAIN}}$$\in \{10, 15, 20,25,30, \inf\}$ dB. The antenna geometry is a UCA with $M=20$ and $K=6$.}
	\label{figSNRTest1DUCAM20SVM}
	\end{figure}

 We also compared CNN with the SVM technique (as in \cite{machineLearningAntennaSelection}) where we used identical parameters for the data generation and identical data covariance input to the SVM. The performance of SVM is shown in  Fig.~\ref{figSNRTest1DUCAM20SVM}. We observe from Figs.~\ref{figSNRTest1DUCAM20}-\ref{figSNRTest1DUCAM20SVM} that CNN is more than $90\%$ accurate for SNR$_{\text{TEST}}\geq10$dB when the network is trained by datasets with SNR$_{\text{TRAIN}} \geq15$dB. In comparison, SVM performs poorly being unable to extract the features as efficiently as CNN. 
	    
\begin{figure}[ht]
		\centering
		{\includegraphics[width=.35\textheight]{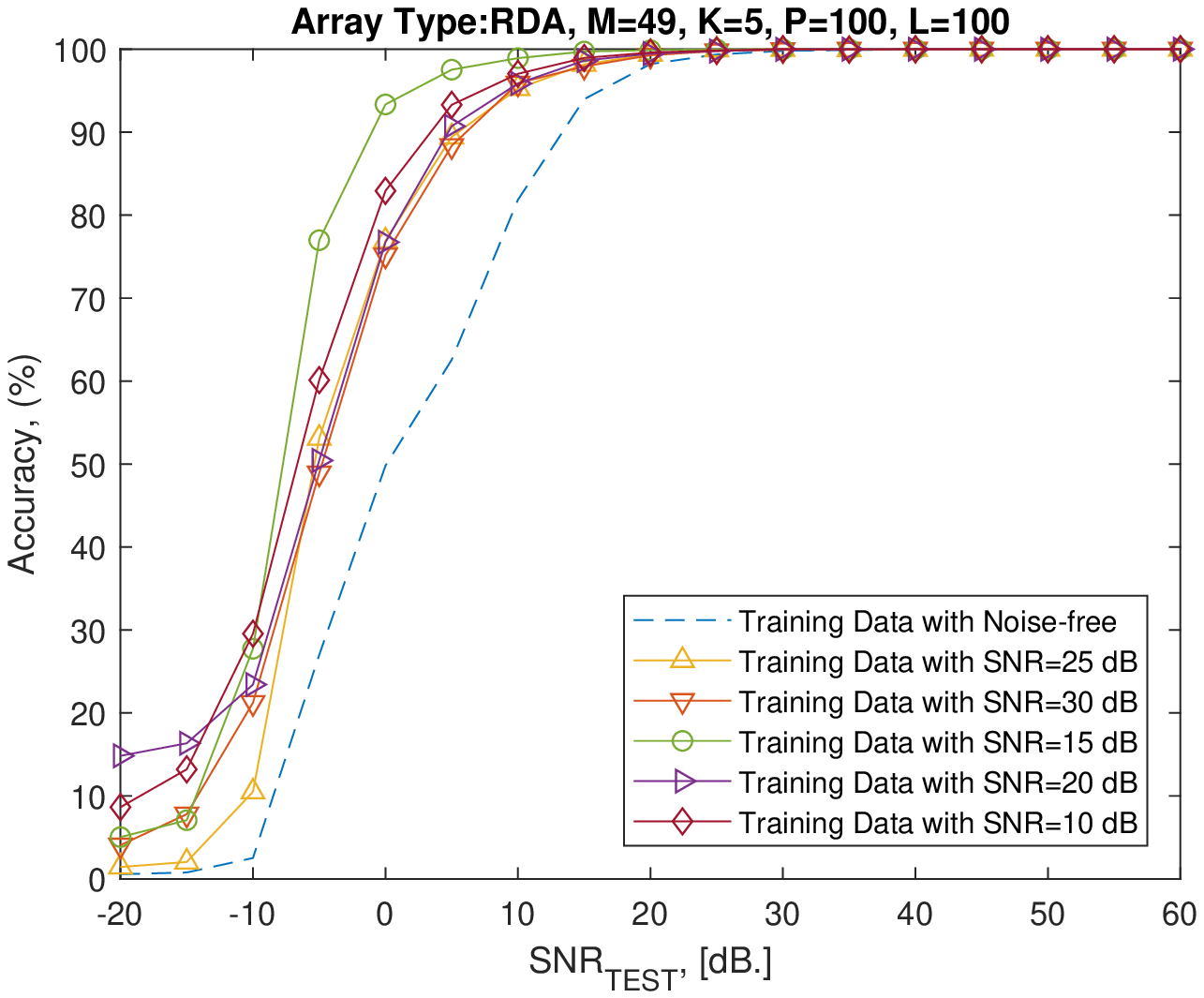}}
		\caption{Performance of test data using CNN with respect to SNR when the training data is prepared with SNR$_{\text{TRAIN}}$ levels of $5$, $10$, $15$, $20$, $25$, and $30$ dB as well as without any noise. The antenna geometry is an RDA with $M=49$ shown in Fig.~\ref{figPosRAM49} and $K=5$.}
		\label{figSNRTest1DRAM49}
	\end{figure}

	Similar experimental results for an RDA (Fig.~\ref{figPosRAM49}) with $M=49$ and $K=5$ are shown in Fig.~\ref{figSNRTest1DRAM49}. The training dataset is prepared with the same parameters as in the previous experiment with UCA. For some selected cases, the accuracies of training and validation data for RDA are listed in Table \ref{tableAccuracies}. The network achieves high accuracy when $M$ is large and SNR$_{\text{TEST}}\geq 10$dB.
	\subsubsection{Experiment \#2: 2-D Scenario}
	Finally, we consider cases when the target and the antenna array are not coplanar. We train the CNN structure in Fig. \ref{figNetwork} with the data generated for $T=100$ and $L=100$. The angles lie on a uniformly spaced elevation and azimuth grids in the planes $[90^{\circ},100^{\circ}]$ and $[0^{\circ},360^{\circ})$, respectively. We set the number of grid points in the elevation and azimuth to $P_{\theta}=11$ and $P_{\phi}=100$, respectively. 
    
    In this experiment, we use an RDA (Fig.~\ref{figPosRAM16}) with $M=16$ and $K=6$. Table~\ref{tableSelectedAntennaIndexes2DScenario} lists the indices of a few RDA antennas that yield the best CRB for different target DoAs: $\phi \in \{30^{\circ},60 ^{\circ},90^{\circ},120 ^{\circ}, 210 ^{\circ}\}$ and $\theta \in \{ 90^{\circ},92^{\circ} \}$ as $K$ varies. When $K$ increases, a subarray with larger aperture is to be selected for better DoA estimation performance. When there is even a slight change in the elevation angle, the best subarray changes completely because of the relatively small subarray aperture in the elevation dimension. We prepared the training and validation datasets for different SNR$_{\text{TRAIN}}$ values. The accuracies of the two stages are listed in Table~\ref{tableAccuracies}. We note that the training accuracy of the network in the 2-D case is worse than the 1-D scenario of RDA because simple 2-D arrays are unable to distinguish all elevation angles. As a result, training data samples that are very similar to each other are labeled to different classes with different elevation angles.

\begin{figure}[ht]
		\centering
		{\includegraphics[width=.35\textheight]{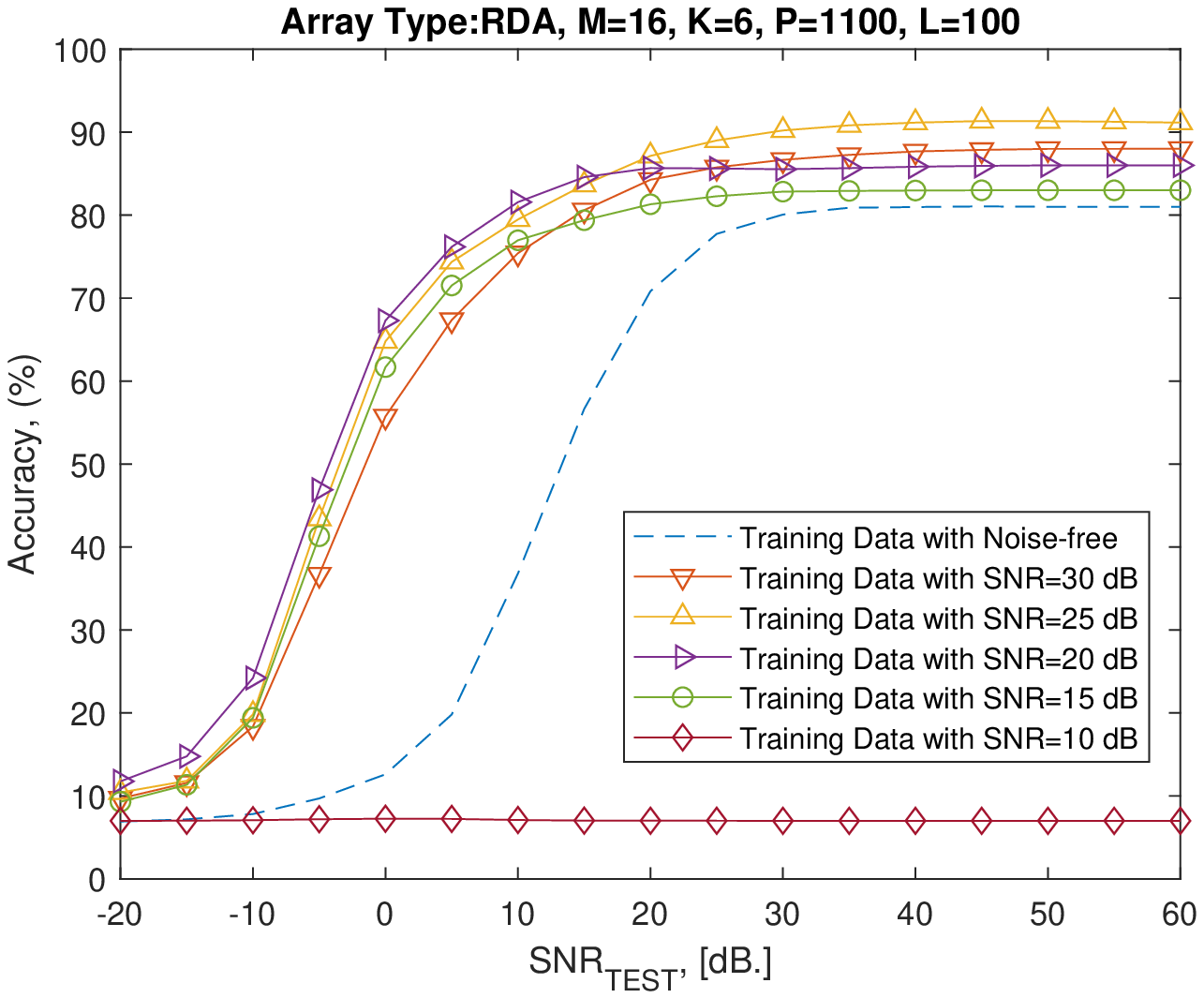}}
		\caption{Performance of test data using CNN with respect to SNR when the training data is prepared with SNR$_{\text{TRAIN}}$ levels of $5$, $10$, $15$, $20$, $25$, and $30$ dB, as well as without any noise. The antenna geometry is an RDA with $M=16$ shown in Fig. \ref{figPosRAM16} and $K=6$.}
		\label{figSNRTest2DScenario}
	\end{figure}

    We generated a test dataset with $L_{\text{TEST}}=100$ and $T_{\text{TEST}}=100$ to evaluate the CNN for a 2-D target scenario. The target directions were drawn from $\phi_{\text{TEST}} \sim \text{u}\{[0^{\circ}, 360^{\circ}]\}$ and $\theta_{\text{TEST}} \sim \text{u}\{[80^{\circ}, 100^{\circ}]\}$ for $P_{\text{TEST}}=100$. 
The accuracies of the test mode for $J_{\text{TEST}}=100$ trials is shown in Fig.~\ref{figSNRTest2DScenario} for different SNR$_{\text{TEST}}$ levels. Figure~\ref{figSNRTest2DScenario} shows that the training datasets with SNR$_{\text{TRAIN}}\geq15$dB provide sufficiently good performance with an accuracy of approximately $85\%$ for SNR$_{\text{TEST}} \geq10$dB. However, as seen earlier, poor classification performance results when SNR$_{\text{TRAIN}}$ is low (e.g., $\leq 10$dB). 
    \begin{figure}[t]
		\centering
		{\includegraphics[width=.35\textheight]{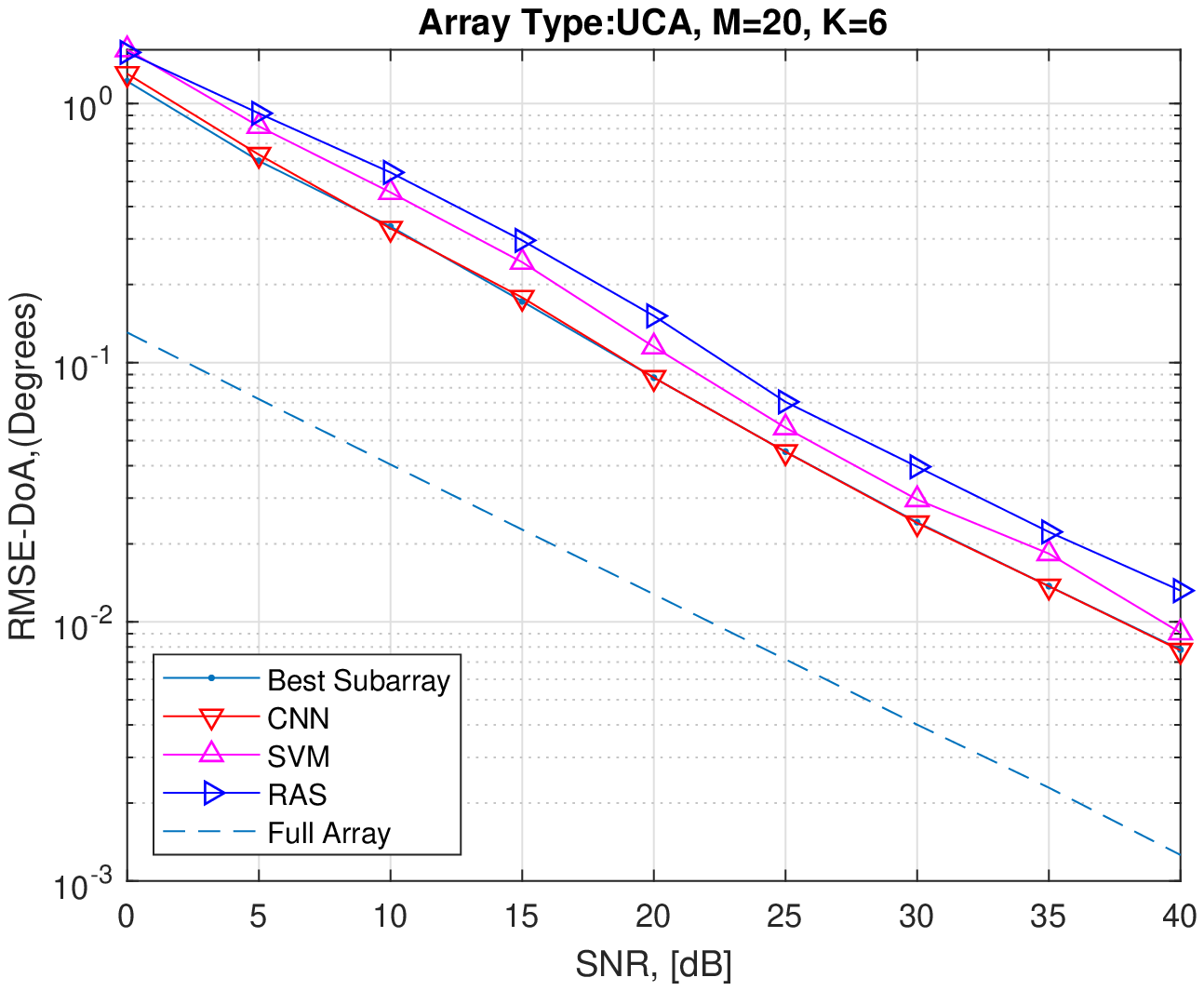}}
		\caption{{DoA estimation performance with respect to SNR. SNR$_{\text{TRAIN}}=20$ dB. The antenna geometry is a UCA with $M=20$ and $K=6$.}}
		\label{figDOAEstimationSNRTestUCA}
	\end{figure}
    \begin{figure}[t]
		\centering
		{\includegraphics[width=.35\textheight]{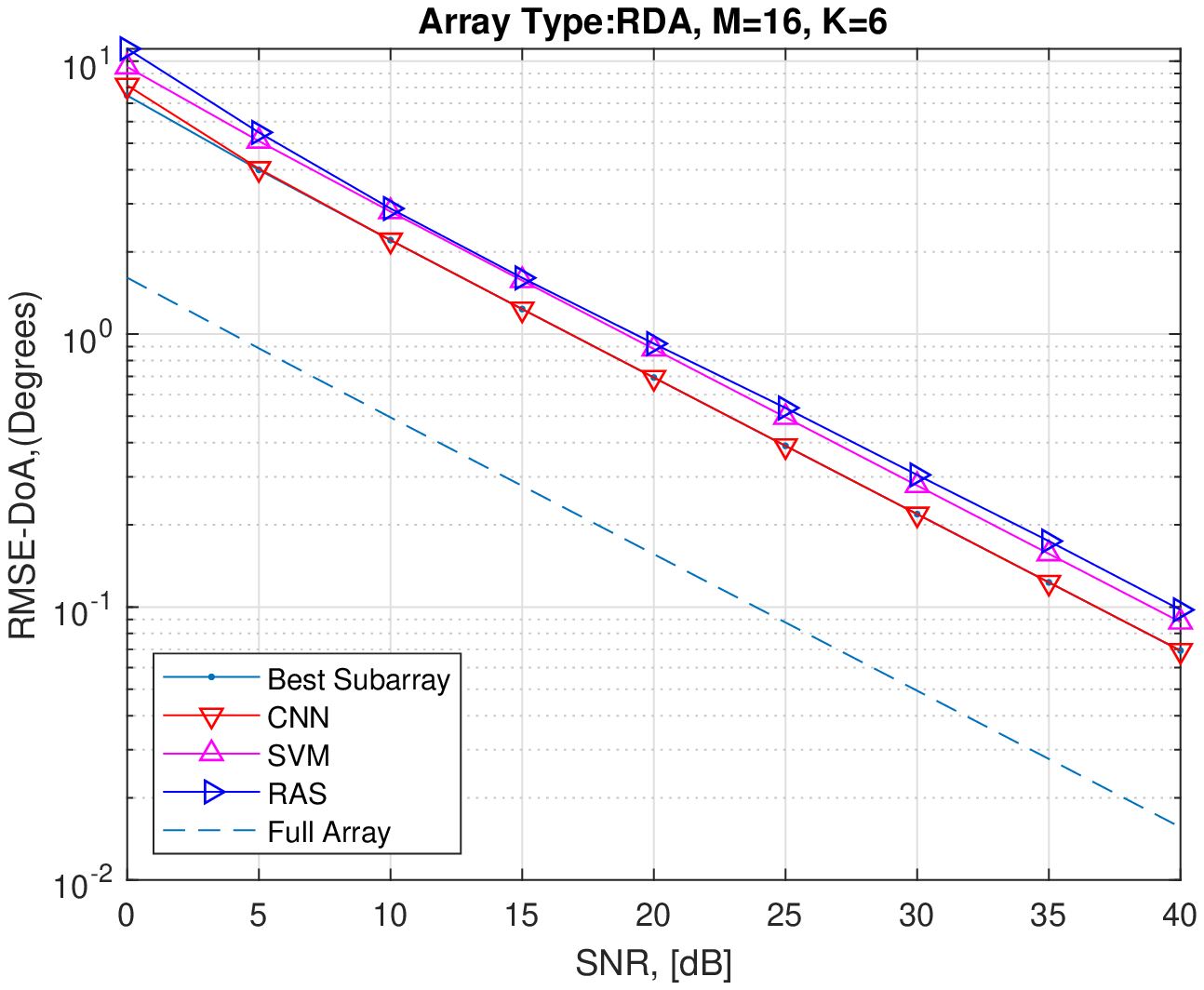}}
		\caption{{DoA estimation performance with respect to SNR. SNR$_{\text{TRAIN}}=20$ dB. The antenna geometry is a RDA with $M=16$ and $K=6$.}}
		\label{figDOAEstimationSNRTestRDA}
	\end{figure}
\begin{figure}[t]
		\centering
		{\includegraphics[width=.35\textheight]{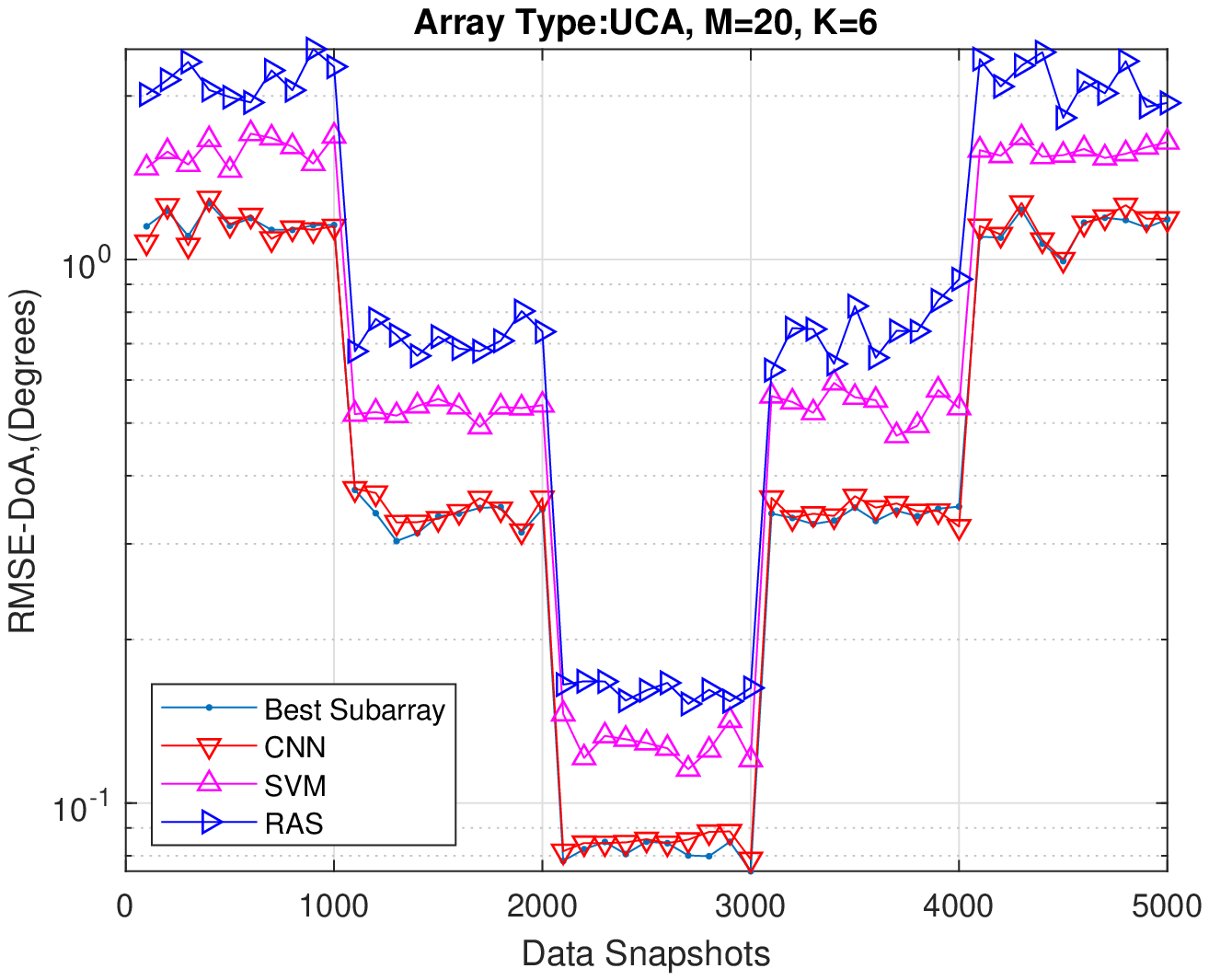}}
		\caption{{DoA estimation performance over time with SNR$_{\text{TRAIN}}=20$ dB. The antenna geometry is a UCA with $M=20$ and $K=6$. SNR$_{\text{TEST}}$ changes over time as $[0, 10, 20, 10, 0]^T$dB.   }   }
		\label{figDOAEstimationOverTimeTestUCA}
	\end{figure}

	{
    \subsubsection{Experiment \#3: DoA Estimation Performance} In this experiment, DoA estimation performance of the proposed method is presented. Our CNN approach is compared with SVM and random antenna selection (RAS) algorithms. The selected antenna subarrays from CNN and SVM are inserted to the beamforming technique \cite{beamformingFundamentals} for DoA estimation. As a traditional technique, we consider the RAS algorithm where, instead of all subarray candidates, a number of subarray geometries are realized randomly (i.e., 1000 realizations) and their beamforming spectra is obtained by a search algorithm \cite{randomArraySelection1}. We also added the full array performance where $M=K$ for comparison. In Fig.~\ref{figDOAEstimationSNRTestUCA}, the results are given for a UCA with $M=20$ and $K=6$ antennas to be selected. Here, "best subarray" denotes the beamforming performance of the subarray that gives the lowest CRB. It can be seen that CNN provides better performance as compared to SVM (32\% more accurate) and RAS (72\%) and it approaches the performance of the "best subarray" as expected from the accuracy results given in Fig.~\ref{figSNRTest1DUCAM20}. SVM performs poorer due to its lower antenna selection accuracy. We present 2D DoA estimation results in Fig.~\ref{figDOAEstimationSNRTestRDA} for RDA with $M=16$ and $K=6$ with the same settings as for Fig.~\ref{figSNRTest2DScenario}. Similar observations are obtained for the 2D case as compared to the 1D scenario.}
     
    {We further compare the DoA estimation performance of the selected subarrays with full array ($M=K$) performance in both Fig.~\ref{figDOAEstimationSNRTestUCA} and Fig.~\ref{figDOAEstimationSNRTestRDA}. While there is a gap between subarray and the full array performances, antenna selection provides less computation and cost.}
     
    {
    In Fig.~\ref{figDOAEstimationOverTimeTestUCA}, the DoA estimation performance over time is presented for different antenna selection algorithms. In this test, SNR is firstly increased from 0dB to 20dB then dropped by 10dB for every 1000 data snapshot blocks. The target location varies for each 500 data snapshot blocks. In each block, the first 100 snapshots are used for antenna selection (all antennas are used). After antenna selection, the selected antennas are used for DoA estimation (only $K$ antennas are in use) for the next data blocks. While the algorithms have robust performance to the change in the target DoA, the CNN has the lowest RMSE as compared to the others.
    }

    \subsection{Computational Complexity}
     The computation times for the algorithms are given in Table~\ref{tableComputationTime} in seconds. In order to fairly compare the algorithms the results are calculated to include both antenna selection and DoA estimation phases. The computation time only for the beamforming is 0.0384s. In terms of classification, the computation time for CNN and SVM are $0.0037$s and $0.0609$s respectively for $M=20$ and $K=6$. As a result, CNN provides much faster results and accuracy as compared to both SVM and the conventional DoA estimation technique based on beamforming. The complexity of RAS is due to the computation of the DoA spectra for each subarray realization (1000 realizations were used for RAS in Table~\ref{tableComputationTime}). {We also compared the computation time for DoA estimation with full array and the CNN with $K$ antennas. We observed that DoA estimation with full array took $0.14$s whereas CNN takes $0.0535$s of which $0.0035$s used for classification. These results show that the proposed CNN approach provides less computational complexity together with the loss in the DoA estimation performance due to the use of less number of antennas as compared to the full array.}

	\section{Discussion and Summary}
	\label{sec:Conclusions}
	We introduced a method based on DL to select antennas in a cognitive radar scenario. We constructed a deep neural network with convolutional layers as a multi-class classification framework. The training data was generated such that each class indicated an antenna subarray that provides the lowest minimal error bound for estimating target DoA in a given scenario. Our learning network then cognitively determines a new array whenever the radar receiver acquires echoes from the target scene. We evaluated the performance of the proposed approach for both 1-D (azimuth) and 2-D (azimuth and elevation) target scenarios using ULA, UCA and RDA structures. The results show enhanced performance of the proposed network over conventional randomly thinned arrays as well as the traditional SVM-based selection. Our method does not depend on the array geometry and selects optimal antenna subarrays for DoA estimation. {As with other classifiers such as SVM,} the classification accuracy of our CNN degrades in low SNR conditions because the network cannot distinguish the array data of different classes and, consequently, predicts false results. We were able to partially mitigate this issue by training the network with noisy data samples. {We design the training data with both single and multiple SNR levels of noisy data to investigate further the performance. The results show that there is a slight difference in the classification accuracy after including data at multiple SNRs during the training. The proposed CNN structure provides 32\% better classification than a Support Vector Machine (SVM) and the resulting subarrays yield 72\% more accurate DoA estimate than a random array selection. The combined computation time required by CNN for the antenna selection and DoA estimation was half of that taken by SVM and three orders of magnitude smaller than a random selection.}
        
		Although the CNN predicts an optimal subarray for 1-D scenarios very well, its performance degrades for 2-D cases. This is expected because the simple 2-D arrays we considered are unable to distinguish all elevation angles and thereby lead to some misclassification. We reserve further investigations of this issue for the future. 

    \begin{table}[t]
		\processtable{{Computation time in seconds. The results include antenna selection and DoA estimation complexity. }\label{tableComputationTime}}
		{\begin{tabular}{|p{2cm}|p{1.5cm}|p{1.5cm}|p{1.5cm}|}
			\hline
			\hline
            &CNN & SVM & RAS\\
            \hline 
            Computation time&0.0414s& 0.0984s & 28.7862s\\
            \hline

		\end{tabular}}
		
	\end{table}
 \section*{Acknowledgment}
K.V.M. and Y.C.E. were funded from the European Union's Horizon 2020 research and innovation programme under grant agreement No. 646804-ERC-COG-BNYQ. K.V.M. also acknowledges partial support via Andrew and Erna Finci Viterbi Fellowship and Lady Davis Fellowship.	

	
	\bibliographystyle{ieeetr}
	\bibliography{main}

\begin{thebibliography}{10}

\bibitem{cognitiveRadarHaykin}
S.~Haykin, ``Cognitive radar: {A} way of the future,'' {\em IEEE Signal
  Processing Magazine}, vol.~23, no.~1, pp.~30--40, 2006.

\bibitem{cognitiveRadarRef2}
J.~R. Guerci, ``Cognitive radar: A knowledge-aided fully adaptive approach,''
  in {\em IEEE Radar Conference}, pp.~1365--1370, 2010.

\bibitem{cognitiveRadarExperiments}
G.~E. Smith, Z.~Cammenga, A.~Mitchell, K.~L. Bell, J.~Johnson, M.~Rangaswamy,
  and C.~Baker, ``Experiments with cognitive radar,'' {\em IEEE Aerospace and
  Electronic Systems Magazine}, vol.~31, no.~12, pp.~34--46, 2016.

\bibitem{cognitiveRadarWaveformDesign}
P.~Chen and L.~Wu, ``Waveform design for multiple extended targets in
  temporally correlated cognitive radar system,'' {\em IET Radar, Sonar \&
  Navigation}, vol.~10, no.~2, pp.~398--410, 2016.

\bibitem{cognitiveRadarWaveformDesignReceiverSelection}
M.~B. Kilani, Y.~Nijsure, G.~Gagnon, G.~Kaddoum, and F.~Gagnon, ``Cognitive
  waveform and receiver selection mechanism for multistatic radar,'' {\em IET
  Radar, Sonar \& Navigation}, vol.~10, no.~2, pp.~417--425, 2016.

\bibitem{mishra2017performance}
K.~V. Mishra and Y.~C. Eldar, ``Performance of time delay estimation in a
  cognitive radar,'' in {\em IEEE International Conference on Acoustics, Speech
  and Signal Processing}, pp.~3141--3145, 2017.

\bibitem{mishra2018cognitive}
K.~V. Mishra, Y.~C. Eldar, E.~Shoshan, M.~Namer, and M.~Meltsin, ``A cognitive
  sub-nyquist mimo radar prototype,'' {\em arXiv preprint arXiv:1807.09126},
  2018.

\bibitem{cognitiveRadarTracking}
K.~L. Bell, C.~J. Baker, G.~E. Smith, J.~T. Johnson, and M.~Rangaswamy,
  ``Cognitive radar framework for target detection and tracking,'' {\em IEEE
  Journal of Selected Topics in Signal Processing}, vol.~9, no.~8,
  pp.~1427--1439, 2015.

\bibitem{cognitiveRadarRecognition}
N.~A. Goodman, P.~R. Venkata, and M.~A. Neifeld, ``Adaptive waveform design and
  sequential hypothesis testing for target recognition with active sensors,''
  {\em IEEE Journal of Selected Topics in Signal Processing}, vol.~1, no.~1,
  pp.~105--113, 2007.

\bibitem{cognitiveRadarSpectrumSharing}
P.~Stinco, M.~S. Greco, and F.~Gini, ``Spectrum sensing and sharing for
  cognitive radars,'' {\em IET Radar, Sonar \& Navigation}, vol.~10, no.~3,
  pp.~595--602, 2016.

\bibitem{cohen2017spectrum}
D.~Cohen, K.~V. Mishra, and Y.~C. Eldar, ``Spectrum sharing radar:
  {C}oexistence via {X}ampling,'' {\em IEEE Transactions on Aerospace and
  Electronic Systems}, vol.~29, pp.~1279--1296, 3 2018.

\bibitem{mishra2018sub}
K.~V. Mishra and Y.~C. Eldar, ``Sub-{N}yquist radar: {P}rinciples and
  prototypes,'' {\em arXiv preprint arXiv:1803.01819}, 2018.

\bibitem{cohen2018sub}
D.~Cohen and Y.~C. Eldar, ``Sub-{N}yquist radar systems: {T}emporal, spectral
  and spatial compression,'' {\em IEEE Signal Processing Magazine}, 2018.
\newblock in press.

\bibitem{na2018tendsur}
S.~Na, K.~V. Mishra, Y.~Liu, Y.~C. Eldar, and X.~Wang, ``{TenDSuR}:
  {T}ensor-based {3D} sub-{N}yquist radar,'' {\em IEEE Signal Processing
  Letters}, 2018.
\newblock in press.

\bibitem{cognitiveRadarReconfigurableCircuitry}
C.~Baylis, M.~Fellows, L.~Cohen, and R.~J.~M. II, ``Solving the spectrum
  crisis: {I}ntelligent, reconfigurable microwave transmitter amplifiers for
  cognitive radar,'' {\em IEEE Microwave Magazine}, vol.~15, no.~5,
  pp.~94--107, 2014.

\bibitem{RPA2}
Y.~Lo, ``A mathematical theory of antenna arrays with randomly spaced
  elements,'' {\em IEEE Transactions on Antennas and Propagation}, vol.~12,
  no.~3, pp.~257--268, 1964.

\bibitem{antennaSelectionForMIMO}
T.~M. Duman and A.~Ghrayeb, ``Antenna selection for {MIMO} systems,'' in {\em
  Coding for MIMO Communication Systems}, pp.~287--315, John Wiley \& Sons,
  2007.

\bibitem{ref_AI4}
A.~Gershman and J.~F. {B\"{o}hme}, ``A note on most favorable array geometries
  for {DOA} estimation and array interpolation,'' {\em IEEE Signal Processing
  Letters}, vol.~4, no.~8, pp.~232--235, 1997.

\bibitem{coprimeDSPConf}
P.~Pal and P.~P. Vaidyanathan, ``Coprime sampling and the {MUSIC} algorithm,''
  in {\em Digital Signal Processing and Signal Processing Education Meeting},
  pp.~289--294, 2011.

\bibitem{nestedArray}
P.~Pal and P.~P. Vaidyanathan, ``Nested arrays: {A} novel approach to array
  processing with enhanced degrees of freedom,'' {\em IEEE Transactions on
  Signal Processing}, vol.~58, no.~8, pp.~4167--4181, 2010.

\bibitem{thinnedArray}
K.~V. Mishra, I.~Kahane, A.~Kaufmann, and Y.~C. Eldar, ``High spatial
  resolution radar using thinned arrays,'' in {\em IEEE Radar Conference},
  pp.~1119--1124, 2017.

\bibitem{rossi2014spatial}
M.~Rossi, A.~M. Haimovich, and Y.~C. Eldar, ``Spatial compressive sensing for
  {MIMO} radar,'' {\em IEEE Transactions on Signal Processing}, vol.~62, no.~2,
  pp.~419--430, 2014.

\bibitem{suMMeRPpaper}
D.~Cohen, D.~Cohen, Y.~C. Eldar, and A.~M. Haimovich, ``{SUMMeR: Sub-Nyquist
  MIMO Radar},'' {\em IEEE Transactions on Signal Processing}, vol.~66,
  pp.~4315--4330, Aug 2018.

\bibitem{antennaSelectionCognitive}
J.~Tabrikian, O.~Isaacs, and I.~Bilik, ``Cognitive antenna selection for {DOA}
  estimation in automotive radar,'' in {\em IEEE Radar Conference}, pp.~1--5,
  2016.

\bibitem{antennaSelectionCognitive2}
O.~Isaacs, J.~Tabrikian, and I.~Bilik, ``Cognitive antenna selection for
  optimal source localization,'' in {\em IEEE International Workshop on
  Computational Advances in Multi-Sensor Adaptive Processing}, pp.~341--344,
  2015.

\bibitem{mateos2017adaptive}
D.~Mateos-N{\'u}{\~n}ez, M.~A. Gonz{\'a}lez-Huici, R.~Simoni, and
  S.~Br{\"u}ggenwirth, ``Adaptive channel selection for {DOA} estimation in
  {MIMO} radar,'' in {\em International Workshop on Computational Advances in
  Multi-Sensor Adaptive Processing}, 2017.
\newblock in press.

\bibitem{antennaSelectionMultipleWavelengthSensing}
G.~Shulkind, S.~Jegelka, and G.~W. Wornell, ``Multiple wavelength sensing array
  design,'' in {\em IEEE International Conference on Acoustics, Speech and
  Signal Processing}, pp.~3424--3428, 2017.

\bibitem{antennaSelectionRxTxPairSelection}
H.~Nosrati, E.~Aboutanios, and D.~B. Smith, ``Receiver-transmitter pair
  selection in {MIMO} phased array radar,'' in {\em IEEE International
  Conference on Acoustics, Speech and Signal Processing}, pp.~3206--3210, 2017.

\bibitem{commPaperMassiveMIMO}
A.~F. Molisch, V.~V. Ratnam, S.~Han, Z.~Li, S.~L.~H. Nguyen, L.~Li, and
  K.~Haneda, ``Hybrid beamforming for massive {MIMO}: {A} survey,'' {\em IEEE
  Communications Magazine}, vol.~55, no.~9, pp.~134--141, 2017.

\bibitem{antennaSelectionTxMIMO}
X.~Zhai, Y.~Cai, Q.~Shi, M.~Zhao, G.~Y. Li, and B.~Champagne, ``Joint
  transceiver design with antenna selection for large-scale {MU-MIMO mmWave}
  systems,'' {\em IEEE Journal on Selected Areas in Communications}, vol.~35,
  no.~9, pp.~2085--2096, 2017.

\bibitem{antennaSelectionMISO}
Z.~Wang and L.~Vandendorpe, ``Antenna selection for energy efficient {MISO}
  systems,'' {\em IEEE Communications Letters}, 2017.
\newblock in press.

\bibitem{antennaSelectionMulticasting}
O.~T. Demir and T.~E. Tuncer, ``Antenna selection and hybrid beamforming for
  simultaneous wireless information and power transfer in multi-group
  multicasting systems,'' {\em IEEE Transactions on Wireless Communications},
  vol.~15, no.~10, pp.~6948--6962, 2016.

\bibitem{antennaSelectionViaCO}
S.~Joshi and S.~Boyd, ``Sensor selection via convex optimization,'' {\em IEEE
  Transactions on Signal Processing}, vol.~57, no.~2, pp.~451--462, 2009.

\bibitem{sparsityEnforcingSS}
V.~Roy, S.~P. Chepuri, and G.~Leus, ``Sparsity-enforcing sensor selection for
  {DOA} estimation,'' in {\em IEEE International Workshop on Computational
  Advances in Multi-Sensor Adaptive Processing}, pp.~340--343, 2013.

\bibitem{antennaSelectionKnapsack}
H.~Godrich, A.~P. Petropulu, and H.~V. Poor, ``Sensor selection in distributed
  multiple-radar architectures for localization: {A} knapsack problem
  formulation,'' {\em IEEE Transactions on Signal Processing}, vol.~60, no.~1,
  pp.~247--260, 2012.

\bibitem{machineLearningRadarDetection}
J.~Metcalf, S.~D. Blunt, and B.~Himed, ``A machine learning approach to
  cognitive radar detection,'' in {\em IEEE Radar Conference}, pp.~1405--1411,
  2015.

\bibitem{svmDoAEst1}
Y.~Gao, D.~Hu, Y.~Chen, and Y.~Ma, ``Gridless {1-b DOA} estimation exploiting
  {SVM} approach,'' {\em IEEE Communications Letters}, vol.~21, no.~10,
  pp.~2210--2213, 2017.

\bibitem{svmDoAEst2}
A.~E. Gonnouni, M.~Martinez-Ramon, J.~L. Rojo-Alvarez, G.~Camps-Valls, A.~R.
  Figueiras-Vidal, and C.~G. Christodoulou, ``A support vector machine {MUSIC}
  algorithm,'' {\em IEEE Transactions on Antennas and Propagation}, vol.~60,
  no.~10, pp.~4901--4910, 2012.

\bibitem{machineLearningAntennaSelection}
J.~Joung, ``Machine learning-based antenna selection in wireless
  communications,'' {\em IEEE Communications Letters}, vol.~20, no.~11,
  pp.~2241--2244, 2016.

\bibitem{deepLearningScience}
Y.~Lecun, Y.~Bengio, and G.~Hinton, ``Deep learning,'' {\em Nature}, vol.~521,
  no.~7553, pp.~436--444, 2015.

\bibitem{deppLearningRepresetation}
Y.~Bengio, A.~Courville, and P.~Vincent, ``Representation learning: {A} review
  and new perspectives,'' {\em IEEE Transactions on Pattern Analysis and
  Machine Intelligence}, vol.~35, no.~8, pp.~1798--1828, 2013.

\bibitem{deepLearningRadarRecognition}
C.~Wang, J.~Wang, and X.~Zhang, ``Automatic radar waveform recognition based on
  time-frequency analysis and convolutional neural network,'' in {\em IEEE
  International Conference on Acoustics, Speech and Signal Processing},
  pp.~2437--2441, 2017.

\bibitem{deepLearning4Radar}
E.~Mason, B.~Yonel, and B.~Yazici, ``Deep learning for radar,'' in {\em IEEE
  Radar Conference}, pp.~1703--1708, 2017.

\bibitem{deepLearningSAR}
C.~P. Schwegmann, W.~Kleynhans, B.~P. Salmon, L.~W. Mdakane, and R.~G.~V.
  Meyer, ``Very deep learning for ship discrimination in {Synthetic Aperture
  Radar} imagery,'' in {\em IEEE International Geoscience and Remote Sensing
  Symposium}, pp.~104--107, 2016.

\bibitem{deepLearnnigRangeDopplerRadar}
B.~Jokanovi\'{c} and M.~Amin, ``Fall detection using deep learning in
  range-{D}oppler radars,'' {\em IEEE Transactions on Aerospace and Electronic
  Systems}, 2017.
\newblock in press.

\bibitem{mishra2018deep}
K.~V. Mishra, A.~Gharanjik, M.~R.~B. Shankar, and B.~Ottersten, ``Deep learning
  framework for precipitation retrievals from communication satellites,'' in
  {\em European Conference on Radar in Meteorology \& Hydrology}, 2018.

\bibitem{performanceBoundsWWB}
A.~Renaux, P.~Forster, P.~Larzabal, C.~D. Richmond, and A.~Nehorai, ``A fresh
  look at the {B}ayesian bounds of the {Weiss-Weinstein} family,'' {\em IEEE
  Transactions on Signal Processing}, vol.~56, no.~11, pp.~5334--5352, 2008.

\bibitem{jiang2001almost}
T.~Jiang, N.~D. Sidiropoulos, and J.~M. ten Berge, ``Almost-sure
  identifiability of multidimensional harmonic retrieval,'' {\em IEEE
  Transactions on Signal Processing}, vol.~49, no.~9, pp.~1849--1859, 2001.

\bibitem{liu2002constant}
X.~Liu and N.~D. Sidiropoulos, ``On constant modulus multidimensional harmonic
  retrieval,'' in {\em IEEE International Conference on Acoustics, Speech, and
  Signal Processing}, vol.~3, pp.~III--2977, 2002.

\bibitem{nion2010tensor}
D.~Nion and N.~D. Sidiropoulos, ``Tensor algebra and multidimensional harmonic
  retrieval in signal processing for {MIMO} radar,'' {\em IEEE Transactions on
  Signal Processing}, vol.~58, no.~11, pp.~5693--5705, 2010.

\bibitem{ben2009lower}
Z.~Ben-Haim and Y.~C. Eldar, ``A lower bound on the {Bayesian MSE} based on the
  optimal bias function,'' {\em IEEE Transactions on Information Theory},
  vol.~55, no.~11, pp.~5179--5196, 2009.

\bibitem{crbStoicaNehorai}
P.~Stoica and A.~Nehorai, ``{MUSIC}, maximum likelihood, and {C}ram\'{e}r-{R}ao
  bound: {F}urther results and comparisons,'' {\em IEEE Transactions on
  Acoustics, Speech, and Signal Processing}, vol.~38, no.~12, pp.~2140--2150,
  1990.

\bibitem{friedlander}
B.~Friedlander and A.~Weiss, ``Direction finding in the presence of mutual
  coupling,'' {\em IEEE Transactions on Antennas and Propagation}, vol.~39,
  no.~3, pp.~273--284, 1991.

\bibitem{ye2008two}
Z.~Ye and C.~Liu, ``{2-D DOA} estimation in the presence of mutual coupling,''
  {\em IEEE Transactions on Antennas and Propagation}, vol.~56, no.~10,
  pp.~3150--3158, 2008.

\bibitem{bibal2017ClassAmbiguity}
A.~Bilal, A.~Jourabloo, M.~Ye, X.~Liu, and L.~Ren, ``Do convolutional neural
  networks learn class hierarchy?,'' {\em IEEE Transactions on Visualization
  and Computer Graphics}, vol.~24, pp.~152--162, Jan 2018.

\bibitem{srivastavaDropoutLayer}
N.~Srivastava, G.~Hinton, A.~Krizhevsky, I.~Sutskever, and R.~Salakhutdinov,
  ``Dropout: {A} simple way to prevent neural networks from overfitting,'' {\em
  Journal of Machine Learning Research}, vol.~15, no.~1, pp.~1929--1958, 2014.

\bibitem{bishop2006pattern}
C.~M. Bishop, {\em Pattern Recognition and Machine Learning}.
\newblock Springer, New York, 2006.

\bibitem{beamformingFundamentals}
D.~E. Dudgeon, ``Fundamentals of digital array processing,'' {\em Proceedings
  of the IEEE}, vol.~65, pp.~898--904, June 1977.

\bibitem{randomArraySelection1}
F.~Athley, ``Optimization of element positions for direction finding with
  sparse arrays,'' in {\em Proceedings of the 11th IEEE Signal Processing
  Workshop on Statistical Signal Processing (Cat. No.01TH8563)}, pp.~516--519,
  Aug 2001.

\end{thebibliography}

\end{document}